\documentclass[journal]{IEEEtran}

\usepackage{caption}
\usepackage{subcaption}
\ifCLASSINFOpdf
  \usepackage[pdftex]{graphicx}
  \graphicspath{{./graph/}}
  \DeclareGraphicsExtensions{.pdf,.jpeg,.png}
\else
  \usepackage[dvips]{graphicx}
  \graphicspath{{./graph/}}
  \DeclareGraphicsExtensions{.eps}
\fi

\ifCLASSINFOpdf
\else
\fi

\usepackage[cmex10]{amsmath,mathtools}
\usepackage{mdwmath}
\usepackage{mdwtab}
\usepackage{amssymb}
\usepackage{xcolor}
\usepackage{algorithmic}
\usepackage{algorithm}
\usepackage{float}

\usepackage{epstopdf}

\usepackage{color}

\usepackage[utf8]{inputenc}

\usepackage{cite}
\usepackage{bm}
\usepackage{hyperref}

\begin{document}

\title{Leverage Variational Graph Representation For Model Poisoning on Federated Learning}

\author{Kai~Li,~\IEEEmembership{Senior Member,~IEEE,}
	Xin~Yuan,~\IEEEmembership{Senior Member,~IEEE,}
        Jingjing~Zheng,~\IEEEmembership{Student Member,~IEEE,}
        Wei~Ni,~\IEEEmembership{Fellow,~IEEE,}
        Falko~Dressler,~\IEEEmembership{Fellow,~IEEE},
        and Abbas Jamalipour,~\IEEEmembership{Fellow,~IEEE}
\thanks{K.~Li is with the Department of Engineering, University of Cambridge, CB3 0FA Cambridge, U.K., and also with Real-Time and Embedded Computing Systems Research Centre (CISTER), Porto 4249--015, Portugal (E-mail: kaili@ieee.org).}
\thanks{J.~Zheng is with CyLab Security and Privacy Institute, Carnegie Mellon University, Pittsburgh, PA 15213, USA, and also with Real-Time and Embedded Computing Systems Research Centre (CISTER), Porto 4249--015, Portugal (E-mail: zheng@isep.ipp.pt).}
\thanks{X.~Yuan and W.~Ni are with the Digital Productivity and Services Flagship, Commonwealth Scientific and Industrial Research Organization (CSIRO), Marsfield, NSW 2122, Australia (E-mail: \{xin.yuan,wei.ni\}@data61.csiro.au).}
\thanks{F.~Dressler is with the School of Electrical Engineering and Computer Science, TU Berlin, Germany (E-mail: dressler@ccs-labs.org).}
\thanks{A.~Jamalipour is with the School of Electrical and Information Engineering, The University of Sydney, Australia (E-mail: a.jamalipour@ieee.org).}
}

\markboth{2023.}%
{Li \MakeLowercase{\textit{et al.}}: Leverage Variational Graph Representation For Model Poisoning on Federated Learning}

\IEEEcompsoctitleabstractindextext{%
\begin{abstract}
\boldmath
This paper puts forth a new training data-untethered model poisoning (MP) attack on federated learning (FL). The new MP attack extends an adversarial variational graph autoencoder (VGAE) to create malicious local models based solely on the benign local models overheard without any access to the training data of FL. Such an advancement leads to the VGAE-MP attack that is not only efficacious but also remains elusive to detection. VGAE-MP attack extracts graph structural correlations among the benign local models and the training data features, adversarially regenerates the graph structure, and generates malicious local models using the adversarial graph structure and benign models' features. Moreover, a new attacking algorithm is presented to train the malicious local models using VGAE and sub-gradient descent, while enabling an optimal selection of the benign local models for training the VGAE. Experiments demonstrate a gradual drop in FL accuracy under the proposed VGAE-MP attack and the ineffectiveness of existing defense mechanisms in detecting the attack, posing a severe threat to FL.
\end{abstract}

}

\maketitle

\IEEEdisplaynotcompsoctitleabstractindextext
\IEEEpeerreviewmaketitle

\section{Introduction}
\label{sec_intro}
Federated learning (FL) has attracted significant attention recently, and emerged as a distributed deep learning paradigm. With FL, each user device trains its local  model with its private data to generate local updates sent to the edge server without sharing the device's private data. The edge server then aggregates the local updates to train a global model, which is sent back to the user devices for the next round of FL training. Based on FL, individual data privacy is protected as no private data is shared~\cite{tan2022towards}. 

Despite the fact that FL offers a protective measure for the data privacy of user devices, it remains susceptible to cyber-epidemic attacks. In these attacks, malevolent entities, such as compromised user devices, execute model or data poisoning strategies. These tactics are designed to manipulate the FL process and proliferate across other innocuous user devices~\cite{li2022internet}. Consequently, this leads to the derailment of the training process and a subsequent degradation in the accuracy of the learning outcomes~\cite{lyu2022privacy}. For the model poisoning attacks, the attacker aims to manipulate the hyperparameters of the benign local model. In contrast, data poisoning attacks involve manipulating the training dataset of benign user devices. To launch effective model or data poisoning attacks~\cite{jagielski2018manipulating}, the attackers need to access the knowledge of the dataset used for FL training, which helps to minimize the detectability of malicious local models. 
FL could be manipulated if an attacker launches model poisoning attacks based solely on the benign local and global models overheard without access to the data. Nevertheless, it is challenging for the attacker to achieve effectiveness and undetectability without knowledge of the data. This type of attack is new, has not yet been discussed in the existing literature, and requires further research to develop effective detection and prevention methods. This new attack underscores the importance of securing FL from local and global training threats.

This paper investigates a new adversarial variational graph autoencoder (VGAE)-based model poisoning (VGAE-MP) attack on FL. VGAE-MP is a new data-untethered cyber-epidemic attack, where malicious local models are generated solely based on the benign local models overheard by attackers and the correlation features of the benign local and global models. This attack could be particularly severe in FL systems under wireless settings, due to the broadcast nature of radio. The attacker starts the VGAE-MP attack by overhearing (or eavesdropping on) the transmissions of local model updates from the benign clients in a communication round. The attacker also has the global model that the server shared in the previous communication round. Then, the attacker executes the VGAE-MP model to craftily generate its malicious local model update that, when aggregated, subtly distorts the global model in the current round. 

Specifically, the attacker manipulates its malicious model update to introduce erroneous gradients or patterns. This is done by running the adversarial VGAE to capture the correlation of the benign local models and then regenerate the graph structure to create malicious local models that can effectively compromise the global and benign local models while remaining indistinguishable from the benign local models. Over time, this insidious injection of inaccuracies shifts the global model away from its optimal learning trajectory, leading to a gradual but significant decline in overall FL accuracy. 

Since the user devices possessing large datasets could improve the learning accuracy of FL, the server selects a portion of the collected local models for the global aggregation. Likewise, the VGAE-MP, as a white-box attack, also selects the benign local models in the training of the VGAE. For example, the user device selection at the attacker ensures that the selected local models have sufficient data features for retrieving the correlation in the VGAE, while the generated malicious local model is within proximity to the global model in Euclidean distance. 

The key contributions of this paper are as follows: 
\begin{itemize}
    \item A new data-untethered model poisoning attack, i.e., VGAE-MP, is proposed to manipulate the correlations of multiple data features in the selected benign local models and maintain the genuine data features substantiating the benign local models; 
    \item A new adversarial VGAE, which is trained together with sub-gradient descent to regenerate the correlations of the local models manipulatively while keeping the malicious local models undetectable. 
\item
The proposed VGAE-MP attack is implemented in PyTorch, showing experimentally that VGAE-MP gradually reduces the accuracy of FL and bypasses the detection of existing poisoning defense mechanisms. This attack can propagate across all benign user devices, which leads to an epidemic infection. The source code of the VGAE-MP attack has been released on GitHub.
\end{itemize}

The remaining of this paper is structured as follows. Section~\ref{sec_relatedwork} introduces the background of adversarial attacks against wireless systems and FL. Section~\ref{sec_systems} investigates the FL system model with malicious agents. The proposed VGAE-MP attack is described in Section~\ref{sec_VGAEMP}. Section~\ref{sec_evaluation} discusses the performance analysis. Section~\ref{sec_cond} concludes the paper. Table~\ref{tb_variables} lists the notation used in the paper.

\begin{table}
\centering
\caption{Notation and definition}
\begin{tabular} {|p{1.5cm}|p{5.8cm}|} \hline
\bf{Notation} & \bf{Definition} \\ \hline
$I$ & number of benign devices \\ 
$J$ & number of attackers \\ 
$D_i(t)$ & datasets of the benign device $i$ at the $t$-th communication round \\
$D$ & total datasets of $I$ number of benign devices \\
$D^\prime(t)$ & the claimed data size of the attacker \\
$\pmb{w}_i(t)$ & local model weight parameters of the benign device~$i$ \\
$\pmb{w}_j^\prime(t)$ & training parameters of the malicious model at attacker $j$ \\
$\pmb{w}_G^\prime(t)$ & the global model of FL under attack \\
$\beta_{i,j}^\prime(t)$ & the binary indicator for selecting benign local model weights \\
$\lambda, \rho$ & the Lagrangian dual variables \\
$\tau_i(t)$ & the training delay of $\pmb{w}_i(t)$ at device $i$ \\
$M$ & total number of model parameters in $\pmb{w}_i$ \\
$\pmb{w}_i^m(t)$ & the $m$-th feature in $\pmb{w}_i$ \\
$\pmb{\mathcal{A}}$ & the adjacency matrix formulated by attacker \\
$\pmb{\mathcal{F}}$ & the feature matrix in VGAE of attacker \\
$\pmb{\mathcal{L}}$ & the Laplacian matrix based on $\pmb{\mathcal{A}}$ \\
$\pmb{\mathcal{L}}_k$ & the rank-$k$ SVD approximation of $\pmb{\mathcal{L}}$ \\
$\eta_{\rm loss}$ & the reconstruction loss of the decoder in VGAE \\
$\widehat{\pmb{\mathcal{A}}}$ & the reconstructed adjacency matrix generated at the decoder of attacker \\
$\widehat{\pmb{\mathcal{F}}}$ & the reconstructed feature matrix at attacker \\
\hline
\end{tabular}
\label{tb_variables}
\end{table}

\section{Related Work}
\label{sec_relatedwork}
This section reviews the literature on adversarial attacks and security threats to FL, e.g., model poisoning, data poisoning, inference, and backdoor attacks. 

\begin{figure*}[htb]
\begin{center}
\begin{tabular}{cc}
\includegraphics[width=3.3in]{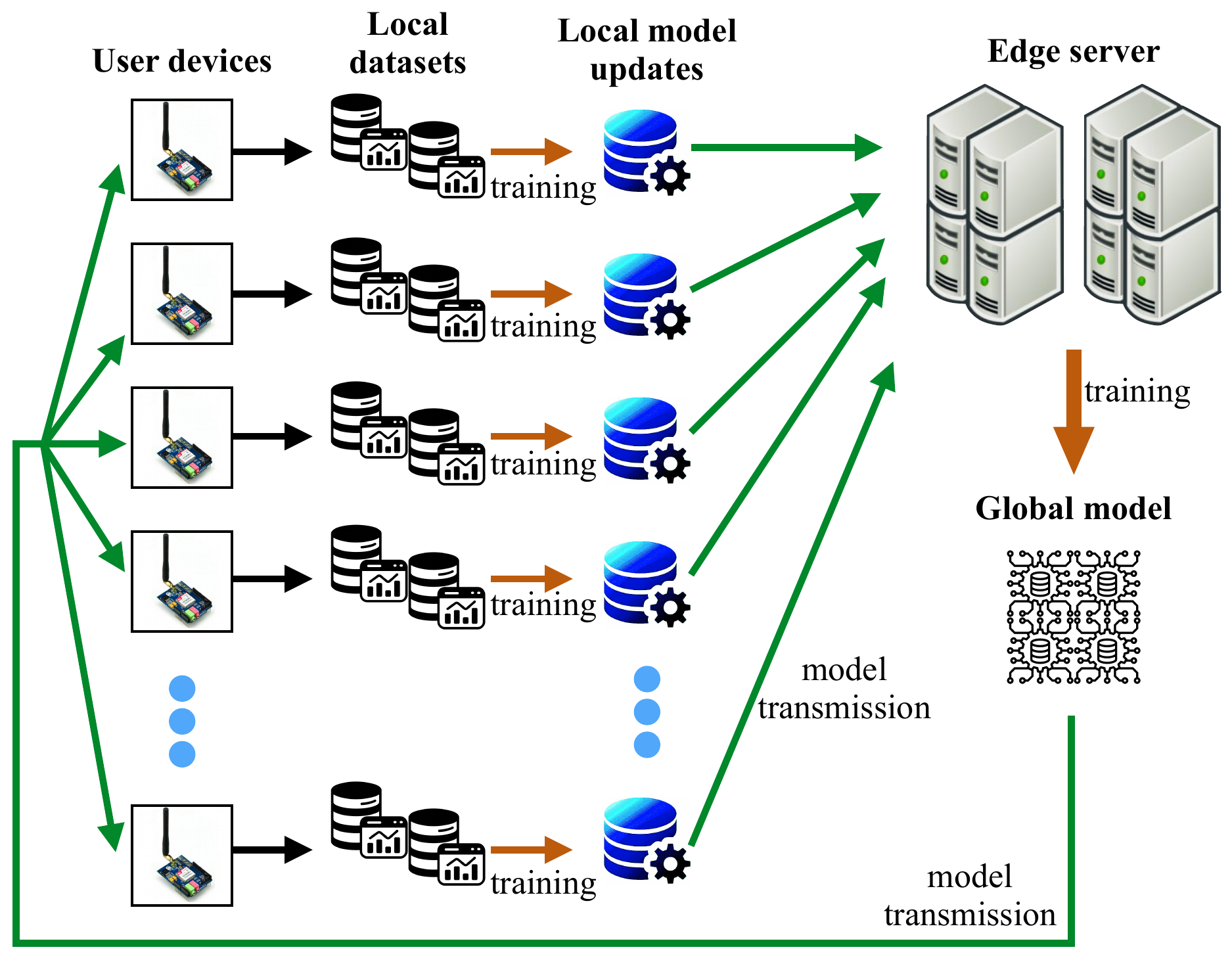} & \includegraphics[width=3.3in]{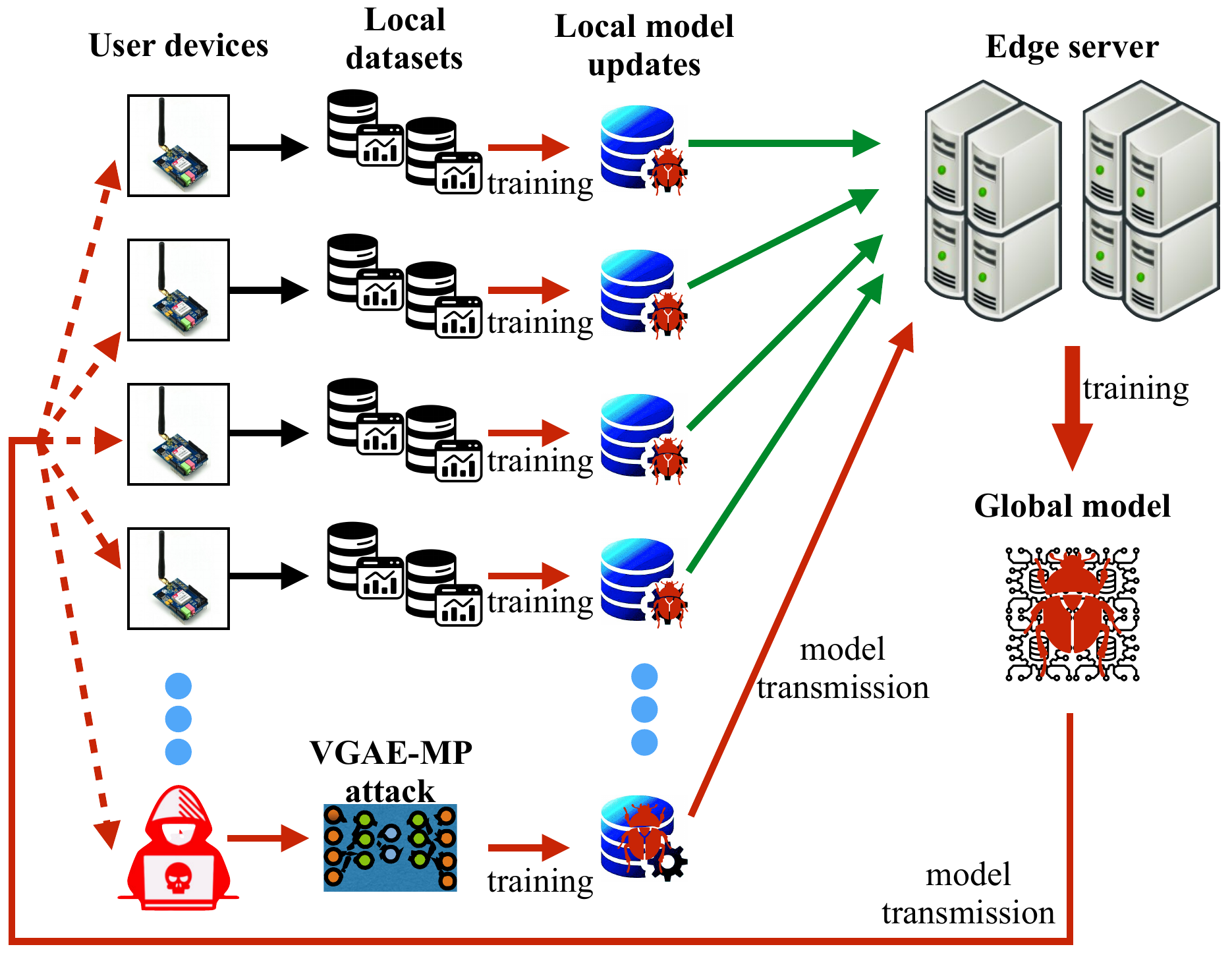} 
\\(a) FL with benign user devices & (b) The proposed VGAE-MP attack
\end{tabular}
\end{center}
\caption{(a) Illustration of FL, where a local model update is trained at each benign user device based on its datasets. The edge server aggregates the benign local model updates to train a global model that will be broadcast to the user devices to update the training parameters of their local models. (b) By eavesdropping on the benign local model updates, the attacker performs the proposed VGAE-MP attack to create a malicious poisoning model that is sent to the server. The malicious model deviates the FL in the opposite direction, thereby falsifying the local model updates of the devices.}
\label{fig_application}
\end{figure*}

The periodic model updates in FL bear a discriminative ability that reflects changes in data distribution, including sensitive properties, making it possible for an attacker to infer unintended information. In~\cite{wang2022poisoning}, the authors introduce a poisoning-assisted property inference attack, which injects malicious data into the training dataset to infer a targeted property of the FL model. The attack modifies the training data labels, thereby distorting the decision boundary of the shared global model in FL, resulting in the disclosure of sensitive property information by benign user devices. 
In~\cite{gao2021secure}, the authors present that the attacker can infer the presence or absence of a particular category in the data by carefully crafting a malicious training dataset, despite the secure aggregation methods. A category inference attack is developed, which iteratively generates malicious training data and utilizes them to update the global model in FL. 
The vulnerability of FL to label inference attacks is presented in~\cite{fu2022label}, where a malicious user device can infer the private labels of other benign devices. With the observed aggregate model updates, three label inference attacks have been developed to infer private labels with the benign devices, including direct, passive, and active label inference attacks.  

In~\cite{gong2022coordinated}, a coordinated backdoor attack on FL is designed using model-dependent triggers, where an attacker can inject a backdoor trigger into a target model and then train the models in FL to perform coordinated attacks. The trigger uses the model dependency in FL to activate the backdoor when the target client uses the compromised model. A distributed attacking algorithm is also provided to enable the attackers to select their respective backdoor models for a high attack success rate while maintaining a low impact on the overall training accuracy of FL.

The authors of~\cite{nuding2022data} focus on data poisoning attacks in both sequential and parallel FL settings. These attacks could weaken the performance of trained models by injecting malicious data into the training datasets used in FL. Sequential FL involves user devices training successively, using the output model from the previous device. In contrast, parallel FL involves each user device simultaneously training a local model before sending updates to the server for aggregation. An attacker can later trigger malicious behavior during the prediction phase by modifying specific training inputs using a specific pattern. In~\cite{fang2020local}, it is argued that FL based on weighted averaging and trimmed averaging for mitigating Byzantine faults is still vulnerable to data poisoning attacks. These attacks can lead to considerable reductions in training accuracy, highlighting a critical vulnerability in the current mitigation strategies within FL. 
A data poisoning attack is studied, which targets the FL system designed to be robust against Byzantine attacks. The data poisoning attack can exploit the characteristics of FL and the Byzantine-robust mechanisms to insert malicious data into the system. 

In~\cite{shejwalkar2021manipulating}, a model poisoning attack is introduced, which accounts for the characteristics of FL, such as variability in the training data and randomness of the training process. The model poisoning attack uses a transfer learning strategy to improve the attack efficiency. 
A model poisoning attack on FL based on fake user devices added to the system and operating as legitimate devices is presented in~\cite{cao2022mpaf}. This fake device can manipulate its data to influence the global model and potentially insert a backdoor or degrade the FL performance. 
In~\cite{chow2021perception}, the authors study a perception poisoning attack in which the attacker manipulates the FL model's perception by altering the training data. The attack can be captured by building a poison perception model for measuring a perception poisoning rate. 

In~\cite{zhang2020poisongan}, the authors focus on a generative poisoning attack against FL, which generates malicious data using generative adversarial networks (GAN) to target user devices in FL. The attack can introduce bias into the aggregated model by injecting poisoned data generated by the GANs. 
Another GAN-based poisoning attack against FL is presented in~\cite{zhang2019poisoning}. The GANs-based poisoning attack creates a set of malicious samples by generating poisoned data samples to attack the local models of benign user devices. To degrade the training accuracy of FL, the attacker deceives the aggregation process at the server by strategically altering the models of the benign user devices. The GAN-based poisoning attack is evaluated based on image classification datasets. 

The existing data or model poisoning attacks against FL lack the description of the implicit relationship between different local models, which can be detected by recent poisoning defense frameworks based on the probabilistic graph model, e.g.,~\cite{li2021lomar,qayyum2022making}. Additionally, convolutional layers at the aggregator can excessively smooth out the output features of the attacks, resulting in distinguishable discrepancies between the malicious local model and the benign ones. In contrast, the proposed VGAE-MP attack is a new attacking method for model poisoning, which is independent of the data. The VGAE-MP attack manipulates the correlations among multiple data features in selected benign local models while preserving the genuine data features that support those models, thus keeping the malicious local models undetectable.

\section{System Model and Problem Statement}
\label{sec_systems}
This section presents the training of the local and global models of FL in mobile edge computing for image classification as an example. Figure~\ref{fig_application}(a) presents an FL training process with $I$ benign user devices. Each benign device $i \in [1, I]$ has $D_i(t)$ data samples at the $t$-th communication round of FL. Let $x(d_i)$ denote a data sample captured at the $i$-th benign device, and $y(d_i)$ the local model update trained at the $i$-th benign device, where $d_i \in [1, D_i(t)]$~\cite{zheng2023federated}. 

The training loss function of a benign device $i$, denoted by  $f(\pmb{w}_i(t); x(d_i), y(d_i))$, measures approximation errors based on the inputs $x(d_i)$ and outputs $y(d_i)$ in the $t$-th communication round, where $\pmb{w}_i(t)\in \mathbb{R}^{1\times M}$ denotes the local model obtained in the communication round. For example, the loss function can be modeled as linear regression, i.e., $f(\pmb{w}_i(t); x(d_i), y(d_i)) = \frac{1}{2} (\pmb{w}_i^T(t) {x(d_i)} - y(d_i))^2$, or logistic regression, i.e., $f(\pmb{w}_i(t); x(d_i), y(d_i)) = y(d_i) \log \Big(1 + \exp \big(-\pmb{w}_i^T(t) {x(d_i)}\big)\Big) - (1- y(d_i)) \log \Big(1-\frac{1}{1 + \exp \big(-\pmb{w}_i^T(t) {x(d_i)} \big)}\Big)$. Here, $(\cdot)^T$ denotes transpose. 

Given $D_i(t)$, the local loss function of the FL at device~$i$ for the $t$-th communication round is 
\begin{equation}
\begin{aligned}
F\!(\pmb{w}_i(t)\!)\!\!=\!\! \frac{1}{D_i(t)}\!\! \!\!\sum_{i=1}^{D_i(t)}&\! f\big(\pmb{w}_i(t);\! x(d_i), \!y(d_i)\!\big) \!\!+\!\! \alpha \zeta\big(\pmb{w}_i(t)\!\big), 
\label{eq_lossFunc}
\end{aligned}   
\end{equation}
where $\zeta(\cdot)$ is a regularizer function capturing the effect of local training noise; $\alpha \in [0,1]$ is a given coefficient~\cite{jiang2022model}.

With the learning rate $\mu$, the local model of device $i$ is updated for $T_L$ local iterations throughout the $t$-th communication round by
\begin{equation}
\pmb{w}_i(t) \leftarrow \pmb{w}_i(t) - \mu \nabla F (\pmb{w}_i(t)),
\label{eq_local_SGD}
\end{equation} 
After the $T_L$ local iterations, all devices upload their local models $\pmb{w}_i(t), \forall i$ to the server. The server aggregates the local models to train a global model denoted by $\pmb{w}_G(t)$ for the $t$-th communication round. Then, $\pmb{w}_G(t)$ is broadcast to all user devices for their training of $\pmb{w}_i(t+1),\,\forall i$ in the $(t+1)$-th communication round. 

Figure~\ref{fig_application}(b) shows the FL of the benign user devices under the proposed VGAE-MP attack, where an attacker overhears $\pmb{w}_i(t)$ uploaded by the benign devices. The attacker, who may appear as a legitimate device, can progressively contaminate the global model represented by $\pmb{w}_G(t)$ and the local models of the benign users, i.e., $\pmb{w}_i(t)$, $\forall i \in [1, I]$, by creating and uploading malicious local models during each communication round $t$. The malicious local model at the attacker's device $j$ is represented by $\pmb{w}_j^\prime(t)$. It is constructed based on the parameters of the benign local models overheard by the attacker during each communication round $t$. 
The server aggregates the local models of the user devices, including both benign and malicious models, without realizing the attacker's presence. This creates a contaminated global model, $\pmb{\omega}_G^\prime(t)$. The total size of the local training data reported to the server, $D(t)$, is calculated as the sum of the data size of all devices, $D_i(t)$, and the claimed data size of the attacker, $D^\prime(t)$. 

Some FL systems allow the server to select a portion of the collected $\pmb{w}_i(t)$ to train $\pmb{w}_G(t)$. For example, the authors of~\cite{zheng2022exploring} considered a selection scheme in which the total bandwidth of the selected devices needs to be smaller than the bandwidth capacity. We define a binary indicator $\beta_{i,j}^\prime(t)$ at the attacker. If $\pmb{w}_i(t)$ is selected by the attacker to train its adversarial and contaminating local model, then $\beta_{i,j}^\prime(t) = 1$; otherwise, $\beta_{i,j}^\prime(t) = 0$. 
Thus, the contaminated global model can be written as
\begin{align}
\pmb{w}_G^\prime(t) =  \sum_{i=1}^I \frac{D_i(t)}{D(t)} \beta_{i,j}^\prime(t) \pmb{w}_i(t) + \frac{D^\prime(t)}{D(t)} \pmb{w}_j^\prime(t),
\label{eq_glbAttacks}
\end{align}
where $\pmb{w}_j^\prime(t)$ is the weight parameter of the malicious model trained at attacker $j$. Then, the server broadcasts $\pmb{w}_G^\prime(t)$ to all $I$ devices. 

\begin{figure*}[htb]
\centering
\includegraphics[width=6.8in]{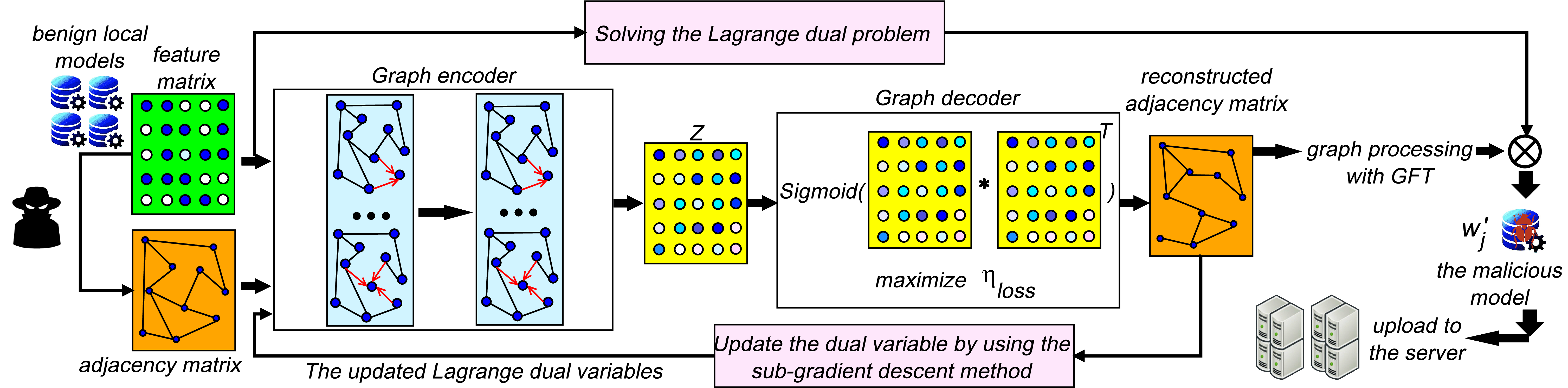}
\caption{The proposed VGAE-MP attack creates $\pmb{w}_j^\prime(t)$ based on learning the correlation among the parameters of the models being trained in FL, i.e., $\pmb{w}_i(t),\,\forall i$. A graph encoder trains $\pmb{\mathcal{F}}_j$ and $\pmb{\mathcal{A}}_j$ to build a feature representation matrix $\pmb{\mathcal{Z}}$. The output of the encoder inputs to the decoder for the reconstruction of $\pmb{\mathcal{A}}_j$. The VGAE-MP attack is designed to adjust $\pmb{w}_j^\prime$ to maximize the reconstruction loss $\eta_{\rm loss}$, according to~\eqref{eq_reconError}.}
\label{fig_graph}
\end{figure*}

The FL trains the global model based on the local datasets of all user devices, including the non-existent dataset claimed by the attacker, by minimizing the global loss function:
\begin{align}
\underset{\pmb w^\prime_G(t)}{\min} F({\pmb w^\prime_G} (t))\!=\!  \sum_{i=1}^I &\frac{D_i(t)}{D(t)} \beta_{i,j}^\prime(t) F_i(\pmb{w}_i(t)) + \nonumber \\ 
&\frac{D^\prime(t)}{D(t)} F_j^\prime(\pmb{w}_j^\prime(t)),
 \label{eq_minGlobalLoss}
\end{align}
where the attacker's claimed local loss function, represented by $F_j^\prime(\cdot)$, is in accordance with~\eqref{eq_lossFunc}.

The optimization of the proposed VGAE-MP attack aims to construct an optimal $\pmb{w}_j^\prime(t)$ based on the overheard $\pmb{w}_i(t)$ to maximize $F({\pmb w^\prime_G} (t))$ in~\eqref{eq_minGlobalLoss}, while maintaining a reasonably small Euclidean distance between $\pmb{w}_j^\prime$ and $\pmb{w}_G^\prime$. This helps $\pmb{w}_j^\prime(t)$ remain undetectable by the server, because the server could evaluate similarities among local models and eliminate those differing significantly using, e.g., Krum or multi-Krum~\cite{blanchard2017machine}. Consequently, ${\pmb w^\prime_G}(t)$ deviates the most in the opposite direction that the benign global model would change in the absence of the attack. 

The optimization of VGAE-MP launched by the attacker~$j$, $\forall j \in [1, J]$, in the communication round $t$ is formulated as 
\begin{subequations}
\begin{align}
~~\max_{\pmb{w}_j^\prime(t), \beta_{i,j}^\prime(t)} & ~~F({\pmb w^\prime_G}(t))\label{eq_opt_obj}\\
{\rm s.t.} &~~d(\pmb{w}_j^\prime(t), {\pmb w^\prime_G}(t)) \leq d_T, \label{eq_const_dist}\\
&~~\sum^I_{i=1} \beta_{i,j}^\prime(t) {d(\pmb{w}_i(t), \Bar{\pmb w}(t))} \leq \Upsilon, \label{eq_const_band}\\
&~~\beta_{i,j}^\prime(t) \in \{0, 1\}, \label{eq_const_select}
\end{align}
\label{eq_opt}
\end{subequations}
where $d(\cdot,\cdot)$ calculates the Euclidean distance between $\pmb{w}_j^\prime$ and $\pmb{w}_G^\prime$, $d_T$ is a threshold of the Euclidean distance, $ \Bar{\pmb w}(t) = \sum_{i=1}^I \frac{D_i(t)}{D(t)} \pmb{w}_i(t)$, and $\Upsilon$ is a predefined upper bound of the overall distance between the selected local models and the aggregated model of the local models.

Constraint~\eqref{eq_const_dist} guarantees that the $j$-th attacker's malicious local model $\pmb{w}_j^\prime$ is in proximity to the global model in terms of Euclidean distance, while constraint~\eqref{eq_const_band} ensures the overall distance between the selected local models and their aggregated model is below the upper bound $\Upsilon$, i.e., $\sum^I_{i=1} \beta_{i,j}^\prime(t)  d(\pmb{w}_i(t), \Bar{\pmb w}(t)) \leq \Upsilon$. This is because the defense mechanism at the FL server, e.g., Krum or multi-Krum, may perform local model selection to rule out those dissimilar to the rest. Constraint~\eqref{eq_const_select} defines $\beta_{i,j}^\prime(t)$ as a binary indicator.

\section{Variational Graph Auto-Encoders-based Model Poisoning Attack on FL}
\label{sec_VGAEMP}
Due to a lack of correlation between the arbitrary features in $\pmb{w}_j^\prime(t)$ and $\pmb{w}_i(t)$, the malicious local model $\pmb{w}_j^\prime(t)$ could be detected by the server. For example, recent graph neural network (GNN)-based FL privacy protection schemes~\cite{wang2022graphfl,yuan2022fedstn} can classify the local model weights based on their features. To tackle this issue, we develop a new adversarial VGAE model in this section to generate $\pmb{w}_j^\prime(t)$ in such a way the individual feature correlation in $\pmb{w}_i(t),\,\forall i$ is captured in $\pmb{w}_j^\prime(t)$. As a result, the server can hardly detect $\pmb{w}_j^\prime(t)$. For the brevity of notation, we omit the subscript $_j$ for the attacker in the following discussions. 

The optimization of VGAE-MP in~\eqref{eq_opt} is a non-convex combinatorial problem intractable for conventional optimization techniques. 
We decouple the VGAE-MP problem in~\eqref{eq_opt} between the model attack and the bandwidth selection using the Lagrangian-dual method~\cite{palomar2006tutorial}. A new approach is developed to iteratively optimize the adversarial local models by running graph autoencoder and subgradient descent, as depicted in Fig.~\ref{fig_graph}.

Let $\lambda$ and $\rho$ denote the dual variables. The Lagrange function of \eqref{eq_opt} is given by 
\begin{equation}
\begin{aligned}\label{eq_lag}
\mathcal{L}(\beta_{i,j}^\prime(t), \lambda, \rho) & = F(\pmb{w}_G^\prime(t)) + \lambda (d_T - d(\pmb{w}_j'(t), \pmb{w}_G^\prime(t))) \\
& \quad + \rho \left(\Upsilon - \sum^I_{i=1} \beta_{i,j}^\prime(t)  {d(\pmb{w}_i(t), \Bar{\pmb w}(t))}\right).
\end{aligned}
\end{equation}
The Lagrange dual function is
\begin{equation}
{\cal D}(\lambda, \rho) = \max_{\pmb{w}_j'(t),\beta_{i,j}^\prime(t)} \mathcal{L}(\beta_{i,j}^\prime(t), \lambda, \rho).
\end{equation}
The dual problem of the primary problem in~\eqref{eq_opt} is 
\begin{equation}\label{eq-dual}
\min_{\lambda, \rho} {\cal D}(\lambda, \rho).
\end{equation}

\subsection{Client Selection}
At communication round $t$, given $\lambda=\lambda(t)$ and $\rho = \rho(t)$, the primary variable $\beta_{i,j}^\prime(t)$ of the bandwidth selection can be optimized by solving
\begin{equation}\label{eq_dual_equation}
\begin{aligned}
\beta_{i,j}^\prime(t)^* = \arg \min_{\beta_{i,j}^\prime(t)} \Bigg\{\sum^I_{i=1} \beta_{i,j}^\prime(t)  {d(\pmb{w}_i(t), \Bar{\pmb w}(t))} \Bigg\},  \,{\rm s.t. } \;\eqref{eq_const_select},  
\end{aligned}
\end{equation}
which is a standard 0/1 knapsack problem and can be readily solved using dynamic programming.

\subsection{Generation of Adversarial Local Models}
A new adversarial VGAE model, leveraging unsupervised learning on graph-structured data according to the variational auto-encoder~\cite{cemgil2020autoencoding}, is proposed to maximize the Lagrange function~\eqref{eq_lag}. 
For given $\beta_{i,j}^\prime(t)^*$, $\lambda(t)$ and $\rho(t)$, we optimize $\pmb{w}_j'(t)$ by solving
\begin{equation}\label{eq_attack_opt}
\begin{aligned}
\pmb{w}_j'(t)^* \!\!= \!\!\!\arg \max_{\pmb{w}_j'(t)} & \bigg\{ \! F(\pmb{w}_G^\prime(t))   \!- \! \lambda(t)  d(\pmb{w}_j'(t), \pmb{w}_G^\prime(t))) \\
& + \rho(t) \sum^I_{i=1} \beta_{i,j}^\prime(t)^*  {d(\pmb{w}_i(t), \Bar{\pmb w}(t))}\bigg\}.    
\end{aligned}
\end{equation}

An attacker positioned within the effective range of the benign device's wireless signal can overhear the transmitted $\pmb{w}_i(t)$ to the server. The level of access an attacker might have depends on the eavesdropping capabilities. For example, standard wireless signals are broadcast in a spherical radius around the transmitting device. This means that the attacker within that radius can have access to the broadcasted signal, where the attacker can capture the traffic and observe the transmitted information~\cite{li2024data}. More advanced attackers might employ methods that allow them to extend the range of their eavesdropping capabilities or to focus on specific directions, allowing them to intercept communications from further away. For example, a highly directional antenna can pick up wireless signals from a much greater distance than a standard antenna.

An attacker, i.e., the $j$-th attacker, can observe the local model parameters of the benign devices to establish the intrinsic correlation between the different parameters of the local models. A graph can be used to characterize the correlation. The graph is then regenerated manipulatively with the VGAE, and used to produce the malicious local model $\pmb{w}_j^\prime(t)$. By this means, we can maximize \eqref{eq_attack_opt} while preventing the convergence of $\pmb{w}_G^\prime(t)$. Constraints~\eqref{eq_const_dist}-\eqref{eq_const_select} are satisfied by designing the decoder of the VGAE to reproduce the correlations. This approach reduces structural dissimilarity between $\pmb{w}_i(t)$ and $\pmb{w}_j^\prime(t)$, which invalidates the existing defense mechanisms. The VGAE is tailored to ensure those constraints and hinder the convergence of the global model by extracting the correlation features between benign local models and embedding the correlation features in graphs for malicious local model generation. 

\subsubsection{Graph Construction and Feature Extraction}

As illustrated in Fig.~\ref{fig_graph}, the graph represented by $\mathcal{G} = ({\cal V}, E, \mathcal{F})$ is utilized to characterize the correlations among the parameters of the models being trained in FL, i.e., $\pmb{w}_i(t),\,\forall i$~\cite{li2023exploring}. The vertexes, edges, and feature matrix of the graph are represented by {${\cal V}$}, $E$, and $\mathcal{F}$, respectively. The VGAE comprises a graph convolutional network (GCN) encoder and an inner product decoder. The encoder encodes the graph data using its features, and the decoder takes the encoded output as input to reconstruct the original graph $\mathcal{G} = ({\cal V}, E, \mathcal{F})$~\cite{wang2020simple}. 

Let $\pmb{\cal{F}}(t) = [\pmb{w}_1(t),\cdots, \pmb{w}_I(t)]^T \in \mathbb{R}^{I \times M}$ be the feature matrix containing all $I$ benign local models at communication round $t$, where $M$ is the dimension of the local model. 
Let $\bm{\omega}^m(t) \in \mathbb{R}^{I \times 1}$ be the $m$-th column of $\pmb{\cal{F}}(t)$.
We use $\delta_{m, m'}(t)$ to denote the \textit{cosine similarity} between the $\bm{\omega}_m (t)$ and $\bm{\omega}_{m'} (t)$ at communication round $t$. $m, m' \in [1, M]$. 
$\delta_{m,m'}(t)$ is defined as~\cite{zhu2020anomaly}
\begin{equation}
\delta_{m,m'}(t) = \frac{(\bm{\omega}^m(t) )^T \bm{\omega}^{m'}(t)}{\|\bm{\omega}^m(t)\| \cdot  \|\bm{\omega}^{m}(t)\|}. 
\label{eq_innerProduct}
\end{equation}
The adjacency matrix, denoted by $\pmb{\mathcal{A}}(t) = [\delta_{m,m'}(t)] \in \mathbb{R}^{M \times M}$, is one of the inputs to the encoder of the VGAE model at the attacker.
According to $\pmb{\mathcal{A}}(t)$, the topological structure of the graph $\mathcal{G}$ can be constructed at the attacker. The feature matrix $\pmb{\mathcal{F}}(t)$ is the other input to the encoder of the VGAE model at the attacker. 

\subsubsection{Encoder design of the VGAE model}
The encoder in the proposed VGAE maps $\mathcal{G}$ to a lower-dimensional representation. We build the encoder based on the GCN architecture, which learns a latent representation that captures the underlying features of $\mathcal{G}$. The encoded representation is then used as input to the decoder to reconstruct the original graph from the lower-dimensional representation to obtain the malicious local model $\pmb{w}_j^\prime(t)$ in~\eqref{eq_dual_equation}.
For the brevity of notation, we omit the index of communication rounds ``$t$'' in the following discussions.

A graph encoder is defined as
\begin{equation}
    \pmb{\mathcal{Z}}_1 = f_{\rm relu} (\pmb{\mathcal{F}}, \pmb{\mathcal{A}}, | \pmb{W}_0);
\end{equation}
\begin{equation}
    \pmb{\mathcal{Z}}_2 = f_{\rm linear} (\pmb{\mathcal{Z}}_1,  \pmb{\mathcal{A}} | \pmb{W}_1),
\end{equation}
where $f_{\rm relu} (\cdot)$ is the Rectified Linear Unit (ReLU) activation function  employed for the first layer, while $f_{\rm linear} (\cdot)$ is the Linear activation function used for the second layer; and
$\pmb{W}_l$ is the learnable parameters specific to the $l$-th layer of the neural networks.

Since determining the probability distribution of the latent representation of vertexes $\pmb{\mathcal{Z}}$ in $\mathcal{G}$ is difficult and intractable~\cite{hasanzadeh2019semi}, we approximate the true posterior by using a Gaussian distribution $\mathcal{N}(\cdot)$, while the encoder takes $\pmb{\mathcal{F}}$ and $\pmb{\mathcal{A}}$ as its input to an inference model parameterized by a two-layer GCN. Thus, we have 
\begin{align}
q(\pmb{\mathcal{Z}} | \pmb{\mathcal{A}}, \pmb{\mathcal{F}}) = \Pi^{M}_{m=1} q(\pmb{z}_m | \pmb{\mathcal{A}}, \pmb{\mathcal{F}}),
\label{eq_Z}
\end{align}
and
\begin{align}
q(\pmb{z}_m | \pmb{\mathcal{A}}, \pmb{\mathcal{F}}) = \mathcal{N}(\pmb{z}_m | \pmb{\mu}_m, {\rm diag}(\pmb{\sigma}^2)),
\end{align}
where $\pmb{\mu} = \pmb{\mathcal{Z}}_2$ builds the matrix of mean vectors $\pmb{\mu}_m$. Likewise, we have $\log \pmb{\sigma} =   f_{\rm linear} (\pmb{\mathcal{Z}}_1,  \pmb{\mathcal{A}} | \pmb{W}_1)$ that shares the first-layer parameters $\pmb{W}_0$.

With the identity matrix $\mathcal{I} \in \mathbb{R}^{M \times M}$, we define $\widetilde{\pmb{\mathcal{A}}} = \pmb{\mathcal{A}} + \mathcal{I}$ with the $(m,m')$-th element $\widetilde{\pmb{\mathcal{A}}}_{m,m'}$,
and the (diagonal) degree matrix ${\pmb{\mathcal{D}}}$ with the $(m,m)$-th element ${\pmb{\mathcal{D}}}_{m,m} =  \sum_{m'=1}^M\widetilde{\pmb{\mathcal{A}}}_{m,m'}$. 
Each layer of the GCN can be written as 
\begin{align}
f_{\mathcal{G}}(\pmb{\mathcal{Z}}_{l-1}, \pmb{\mathcal{A}} | \textbf{W}_l) = \phi({\pmb{\mathcal{D}}}^{-\frac{1}{2}} \widetilde{\pmb{\mathcal{A}}} {\pmb{\mathcal{D}}}^{-\frac{1}{2}} \pmb{\mathcal{Z}}_{l-1} \textbf{W}_l), 
\label{eq_encoder}
\end{align}
where $ \phi(\cdot)$ is the activation function such as ${\rm relu}(\cdot)$. 

\subsubsection{Decoder design of the VGAE model}
The input to the decoder of the proposed VGAE model is $\pmb{\mathcal{Z}}$, which is the output of the GCN in the encoder. The decoder aims to reconstruct $\pmb{\mathcal{A}}$, denoted by $\widehat{\pmb{\mathcal{A}}}$, predicting whether there is a link between two vertexes by an inner product between latent variables, which is designed as 
\begin{align}
p(\widehat{\pmb{\mathcal{A}}} | \pmb{\mathcal{Z}}) = \sum^M_{m=1} \sum^M_{m'=1} p(\hat\delta_{m,m'} | \pmb{z}_m, \pmb{z}_{m'});
\label{eq_decoder}
\end{align}
\begin{align}
p(\hat\delta_{m,m'} = 1 | \pmb{z}_m, \pmb{z}_{m'}) = \text{sigmoid} (\pmb{z}_m^T \pmb{z}_{m'}), 
\end{align}
where 
$\pmb{z}_m \in \mathbb{R}^{M \times 1}$ is the $m$-th column of $\pmb{\mathcal{Z}}$, and \text{sigmoid}($\cdot$) is the logistic sigmoid function, i.e., ${\rm sigmoid} (x) = 1/(1+\exp^{-x})$. 
Here, the larger the inner product $(\pmb{z}_m^T \pmb{z}_{m'})$ in the embedding, the more likely vertexes $m$ and $m'$ are connected in the graph, according to $\widehat{\pmb{\mathcal{A}}} = [\hat\delta_{m,m'}] \in \mathbb{R}^{M \times M}$ in the autoencoder~\cite{pan2019learning}.

We can view~\eqref{eq_decoder} as the inverse operation of the encoder for constructing a reconstructed adjacency matrix $\widehat{\pmb{\mathcal{A}}}$ as the output of the decoder.
A reconstruction loss function $\eta_{\rm loss}$ is defined at the decoder to measure the difference between $\pmb{\mathcal{A}}$ and $\widehat{\pmb{\mathcal{A}}}$. Given~\eqref{eq_Z} and~\eqref{eq_decoder}, $\eta_{\rm loss}$ is given as 
\begin{align}
\eta_{\rm loss}\! = \!\mathbb{E}_{q(\pmb{\!\mathcal{Z}} | \pmb{\mathcal{A}}, \pmb{\mathcal{F}})} \Big[ \!\log p(\widehat{\pmb{\mathcal{A}}} | \pmb{\mathcal{Z}}) \!\Big] \!- \!\Phi[q(\pmb{\mathcal{Z}} | \pmb{\mathcal{A}}, \pmb{\mathcal{F}}) |p(\pmb{\mathcal{Z}})], 
\label{eq_reconError}
\end{align}
where $p(\pmb{\mathcal{Z}}) \!\!= \!\!\Pi_m p(\pmb{z}_m) \!\!= \!\!\Pi_m \mathcal{N}(\pmb{z}_m | 0, \mathcal{I})$ provides a Gaussian prior, and $\Phi[q(\pmb{\mathcal{Z}} | \pmb{\mathcal{A}}, \pmb{\mathcal{F}}) | p(\pmb{\mathcal{Z}})]$ provides the Kullback-Leibler divergence~\cite{joyce2011kullback} between $q(\pmb{\mathcal{Z}} | \pmb{\mathcal{A}}, \pmb{\mathcal{F}})$ and~$p(\pmb{\mathcal{Z}})$. 

\subsubsection{Generation of adversarial local models $\pmb{w}_j^\prime(t)$}
The Laplacian matrix of $\mathcal{G}$~\cite{molitierno2016applications} is built based on the adjacency matrix of the benign models, i.e., $\pmb{\mathcal{A}}$, as given by 
\begin{align}
\mathcal{L} = diag(\pmb{\mathcal{A}}) - \pmb{\mathcal{A}}.
\label{eq_Laplacian}
\end{align}
By applying singular value decomposition (SVD)~\cite{lange2010singular} to $\mathcal{L}$, i.e., $\mathcal{L}=B\Sigma B^T$, we can obtain a complex unitary matrix $B \in \mathbb{R}^{J \times J}$, also known as graph Fourier transform (GFT) basis, that is used to transform graph data, e.g., $\pmb{\mathcal{F}}$, to its spectral-domain representation. $\Sigma\in \mathbb{R}^{J \times J}$ is a diagonal matrix with the eigenvalues of $\mathcal{L}$ along its main diagonal. 

Due to the abundance of local training data at a client, $\pmb{w}_i^m(t)$ typically contains numerous model parameters, i.e., $M\gg 1$, which leads to a large size of $\pmb{\mathcal{A}} = \{\delta_{m,m'}\} \in \mathbb{R}^{M \times M}$. The exact SVD of $\mathcal{L}$ that has an $M \times M$ matrix has time complexity $\mathcal{O}(M^3)$, which is infeasible in the presence of a large $\pmb{\mathcal{A}}$. To reduce the dimensionality of $\pmb{\mathcal{A}}$ while preserving the features, we consider a fast low-rank SVD approximation~\cite{menon2011fast}, which retains the $k$ singular values and their corresponding singular vectors, where $k \ll M^3$. In particular, a truncated SVD of $\mathcal{L}$ can be formulated as $\mathcal{L}_k \approx B_k \Sigma_k B_k^T$, where $B_k$ is an $m \times k$ matrix with columns being the first $k$ left singular vectors of $\mathcal{L}$, $\Sigma_k$ is a $k \times k$ diagonal matrix with entries being the first $k$ singular values of $\mathcal{L}$, and $B_k$ is an $n \times k$ matrix with columns being the first $k$ right singular vectors of $\mathcal{L}$.

With $B$ (or more explicitly, $B_k$), an attacker, i.e., attacker $j$, can obtain a matrix $S$ that contains the spectral-domain data features of all $\bm{\omega}^m(t)$, $\forall m$ by removing the correlations among the models and subsequently focusing on the data features substantiating the local models. $S$ is given by~\cite{SHAN2023108950} 
\begin{align}
S = B^{-1}_k \pmb{\mathcal{F}}.
\label{eq_S}
\end{align}

Likewise, the attacker can produce a Laplacian matrix based on the output of the VGAE, as given by 
\begin{align}
\widehat{\mathcal{L}} = diag(\widehat{\pmb{\mathcal{A}}}) - \widehat{\pmb{\mathcal{A}}}.
\label{eq_maliciousLap}
\end{align}
The corresponding GFT basis, denoted by $\widehat{B}_k$, can be obtained by applying the fast low-rank SVD approximation to $\widehat{L}$. With reference to~\eqref{eq_S}, the malicious local model that follows $\pmb{\mathcal{A}}$ in the VGAE can be determined by
\begin{align}
\widehat{\pmb{\mathcal{F}}} = \widehat{B}_k S, 
\label{eq_maliciousF}
\end{align}
where $\widehat{\pmb{\mathcal{F}}} \in \mathbb{R}^{I \times M}$. The $j$-th row vector of $\widehat{\pmb{\mathcal{F}}}$ is selected as the malicious local model  $\pmb{w}_j^\prime(t)$ and uploaded by the $j$-th attacker to the aggregator for global model aggregation in communication round $t$.

\subsection{Update of Dual Variables}
Given the attack model $\pmb{w}_j'(t)$, with the obtained $\beta_{i,j}^\prime(t)^*$, the sub-gradient descent method can be taken to update $\lambda(t)$ and $\rho(t)$ by solving the dual problem~\eqref{eq-dual}. Specifically, $\lambda(t)$and $\rho(t)$ are updated by~\cite{boyd2004convex}
\begin{subequations}\label{eq-sub-gradient}
\begin{align}
   & \lambda\left(t +1 \right)  \!= \!\left[ \lambda(t ) - \varepsilon\left( d(\pmb{w}_j'(t), \pmb{w}_G^\prime(t)) - d_T \right) \right]^+;\\
   &  \rho\left(t +1 \right)  \!= \!\left[ \rho(t ) \!- \!\varepsilon\left( \sum^I_{i=1} \beta_{i,j}^\prime(t)^*  { d(\pmb{w}_i(t), \Bar{\pmb w}(t))}\! - \!\Upsilon \right) \right]^+,
\end{align}
\end{subequations}
where $\varepsilon$ is the step size, and $\left[x \right]^+ = \max\left(0,x \right)$. At initialization, $\lambda(t)$ and $\rho(t)$ are non-negative, i.e., $\lambda(1) \geq 0$ and $\rho(1) \geq 0$, to ensure \eqref{eq-sub-gradient} converges.  

Since the attacker aims to generate the malicious local models to disorient FL, the proposed VGAE is constructed and trained to maximize $\eta_{\rm loss}$. As a consequence, $\pmb{w}_j^\prime(t)$ progressively and increasingly contaminates the FL training, as global model aggregations increase, i.e., $t=1,2,3,\cdots$.

\subsection{Algorithm Design of The VGAE-MP Attack}
According to the design of the new VGAE-MP attack in Figure~\ref{fig_graph}, Algorithm~\ref{alg_flaph} is developed along with the FL training of the benign user devices and the FL server. Specifically, the FL server broadcasts $\pmb{w}_G^\prime$ in every communication round. Each benign node $i$ ($1\leq i \leq I$) applies the LocalTraining\_start($\pmb{w}_G^\prime$) function for training the local model $\pmb{w}_i$. 
Each attacker, i.e., the $j$-th attacker ($1\leq j \leq J$), overhears the global model $\pmb{w}_G^\prime$ and the local model $\pmb{w}_i$ from the benign nodes. 
The GAE is trained to maximize the reconstruction loss with $\pmb{\mathcal{A}}$ and $\pmb{\mathcal{F}}$. At the output of the GAE, the attacker achieves the optimal malicious local model, i.e., $\pmb{w}_j^\prime$. 
Then, $\pmb{w}_j^\prime$ is uploaded to the FL server for aggregation. As $\pmb{w}_j^\prime$ is highly correlated with $\pmb{w}_i$ from the benign user devices, the FL server is unlikely to detect and identify the attacker. 

\begin{algorithm}[t]
\caption{The proposed VGAE-MP attack algorithm}
\label{alg_flaph}
\begin{algorithmic}[1]
\STATE{\textbf{1. Initialize}: $\mathcal{G} = ({\cal V}, E, \mathcal{F})$, $T_L$, $I$, $J$, $d_T$, $\pmb{w}_G^\prime(t)$, $\pmb{w}_i^m(t)$, and $\lambda(1) \geq 0$.}\\
{\% \textbf{Adversarial FL}:}
\FOR{round $t=1,2,3,\cdots$}
\FOR{Local iteration number $l = 1,\cdots, T_L$}
\STATE{All benign user devices train their benign local model $\pmb{\omega}_i(t)$, $i =1,\cdots, I$.}
\ENDFOR
\STATE{All benign user devices upload their benign local models $\pmb{w}_i(t)$, $i=1,\cdots, I$ to the server, and the attackers overhear the benign local models.}
\STATE{The attacker $j$ carries out the proposed VGAE, i.e., {\textbf{VGAE}}({$\bm{\omega}^m(t), \forall m,\pmb{\mathcal{F}}, \lambda(t)$}), and obtains $\pmb{w}_j^\prime(t)$:}\\
\STATE{\hspace{4 mm} $\boldsymbol{\cdot}$ Build the adjacency matrix $\pmb{\mathcal{A}} = [\delta_{m,m'}] \in$ \\ \hspace{5 mm} $\mathbb{R}^{M \times M}$ according to~\eqref{eq_innerProduct}, and input $\pmb{\mathcal{A}}$ and $\pmb{\mathcal{F}}$ \\ \hspace{5 mm}into the VGAE. }%
\STATE{{\hspace{4 mm} $\boldsymbol{\cdot}$ Train the VGAE to maximize the reconstruction \\ \hspace{4 mm}  loss {$ \eta_{\rm loss}$} to obtain $\widehat{\pmb{\mathcal{A}}}$.}}
\STATE{\hspace{4 mm} $\boldsymbol{\cdot}$ Obtain $S$ based on~\eqref{eq_Laplacian} and~\eqref{eq_S}, next obtain\\
\hspace{4 mm} $ \widehat{\pmb{\mathcal{F}}}$ based on~\eqref{eq_maliciousLap} and~\eqref{eq_maliciousF}, and then determine \\
\hspace{4 mm} $\bm{\omega}^m(t)$ based on~$\widehat{\pmb{\mathcal{F}}}$}.
\STATE{Update {$\lambda(t)$}, according to \eqref{eq-sub-gradient}.}
\STATE{The attacker uploads the malicious local model $\pmb{w}_j^\prime(t)$ to the server.}
\STATE{The server aggregates selected local models to obtain the global model {under attack $\pmb{w}_G^\prime(t)$ by~\eqref{eq_glbAttacks}, and broadcasts $\pmb{w}_G^\prime(t)$}.}
\STATE{All benign user devices update their local models with the global model, i.e., {$\pmb{w}_i(t) \leftarrow \pmb{w}_G^\prime(t),\,\forall i$}.}
\ENDFOR
\end{algorithmic}
\end{algorithm}

Note that an attacker positioned in proximity to benign devices and equipped with radio transceivers has the potential to passively eavesdrop on the transmitted local models from one or more benign devices. This allows the attacker to discern their characteristics and subsequently devise a malicious local model. The more benign local models are overheard, the more profound the exploration into the feature correlation between the benign local and global models, and the more unlikely the malicious local models are detected by the server. The VGAE-MP attack remains operational even if only a single benign local model is overheard, though its effectiveness is diminished compared to scenarios where multiple benign local models are eavesdropped upon. 

Although cryptography can prevent eavesdropping attacks to some extent, existing techniques, such as those developed in~\cite{hebrok2023we} and~\cite{diaz2019tls}, have demonstrated the possibility of deciphering encrypted information with limited initial data. This risk is even more threatening with the rapid advancement of Quantum computing. The proposed data-untethered VGAE-MP attack could potentially work in compiling with these attack techniques to evade cryptographic protection of the benign local models and poison the training of FL.

\section{Performance Evaluation}
\label{sec_evaluation}
This section demonstrates the implementation of the proposed new VGAE-MP attack in PyTorch. Based on MNIST handwritten digits~\cite{deng2012mnist}, FashionMNIST and CIFAR-10 datasets~\cite{xiao2017fashion}, the training accuracy of the local and global models under the attack is evaluated. The detection rate of the VGAE-MP attack is also presented, which is measured according to the Euclidean distance between the malicious local model and the benign one. 
The source code of the proposed VGAE-MP attack is available on GitHub: \url{https://github.com/jjzgeeks/VGAE-based\_Model\_Poisoning\_Attack\_FL}. 

\subsection{Implementation of The VGAE-MP Attack}
\begin{figure*}[!t]
\begin{center}
\begin{tabular}{ccc}
\includegraphics[width=2.2in]{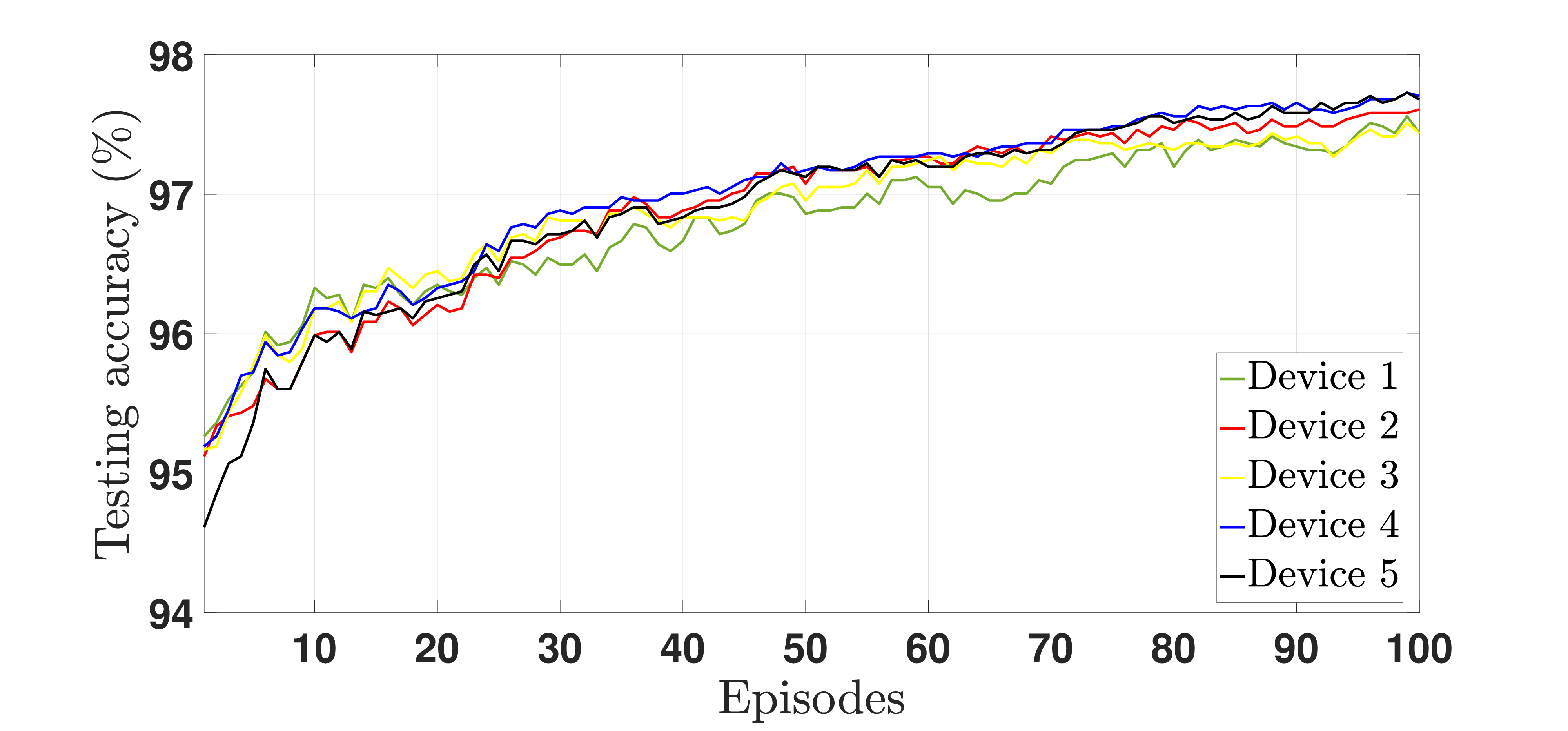} & \includegraphics[width=2.2in]{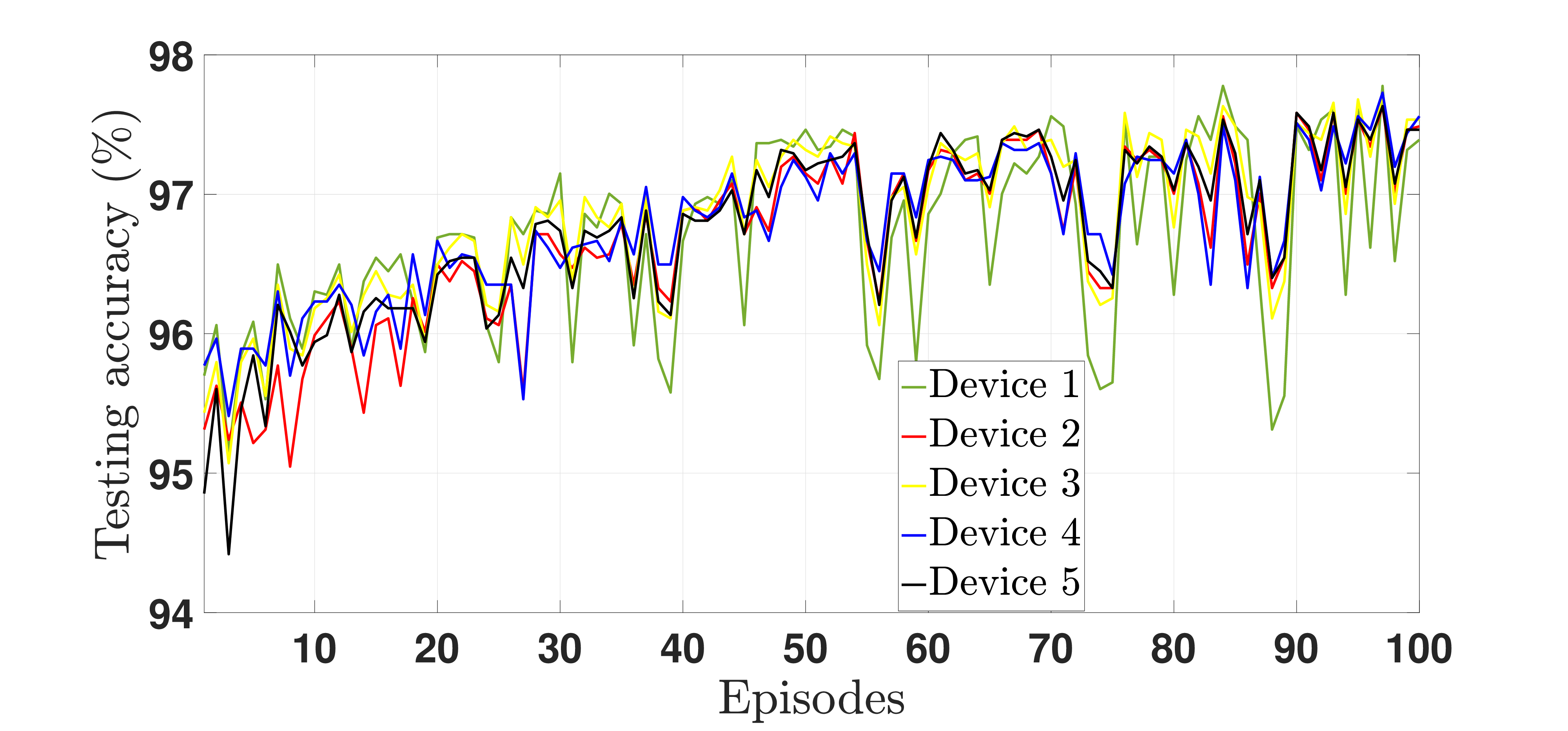} & \includegraphics[width=2.2in]{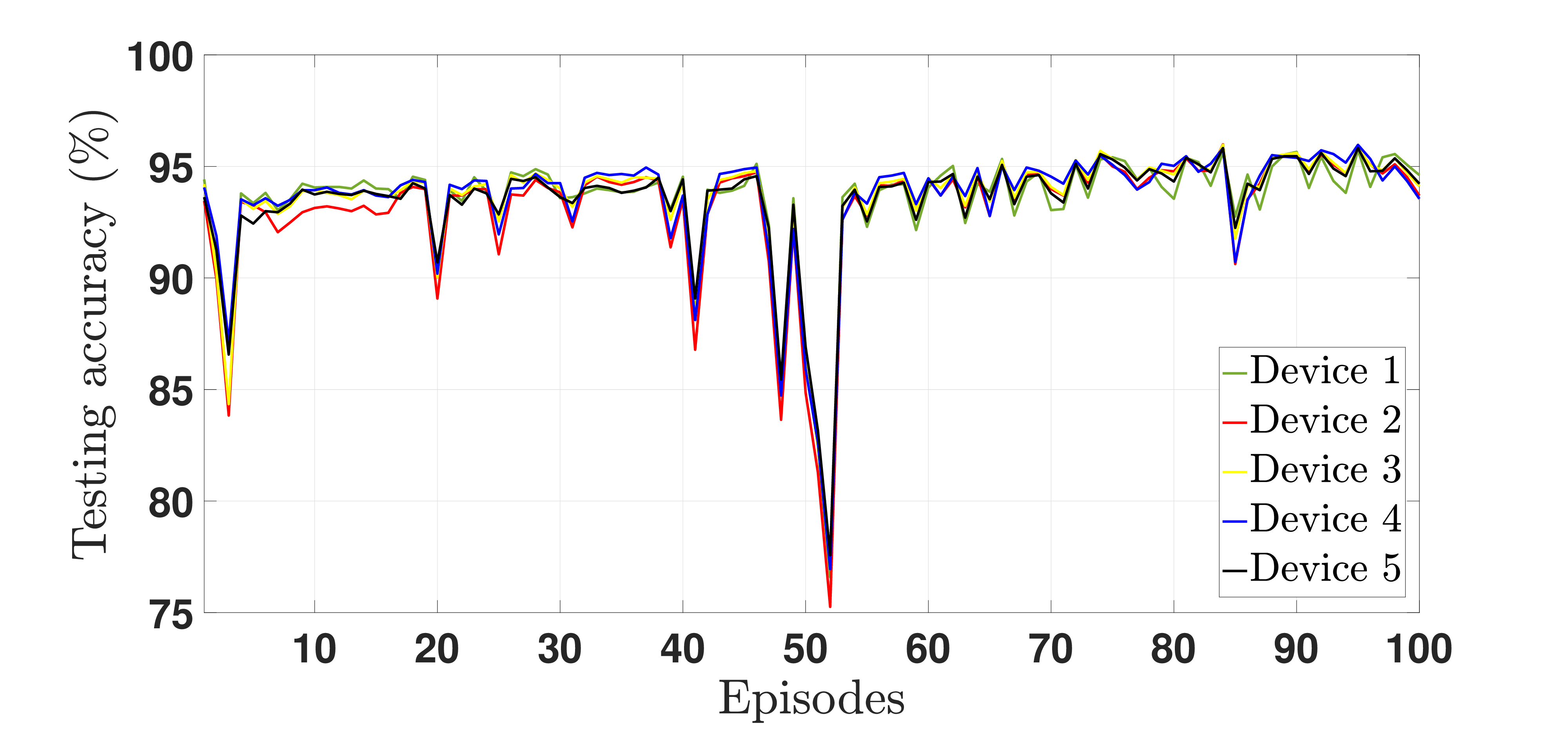}
\\ (a) $M$ = 100 with MNIST. & (b) $M$ = 200 with MNIST. & (c) $M$ = 300 with MNIST.
\\ \includegraphics[width=2.2in]{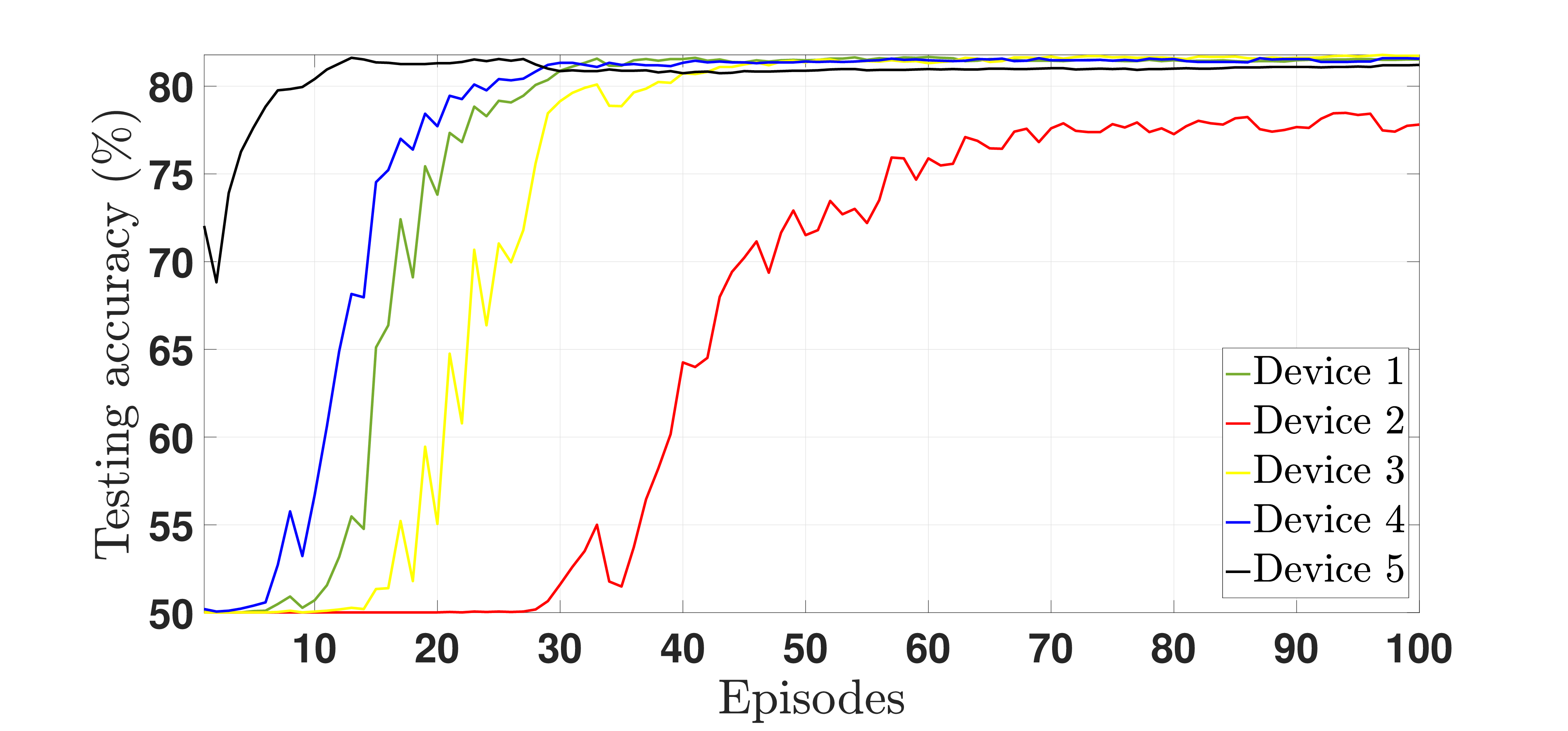} & \includegraphics[width=2.2in]{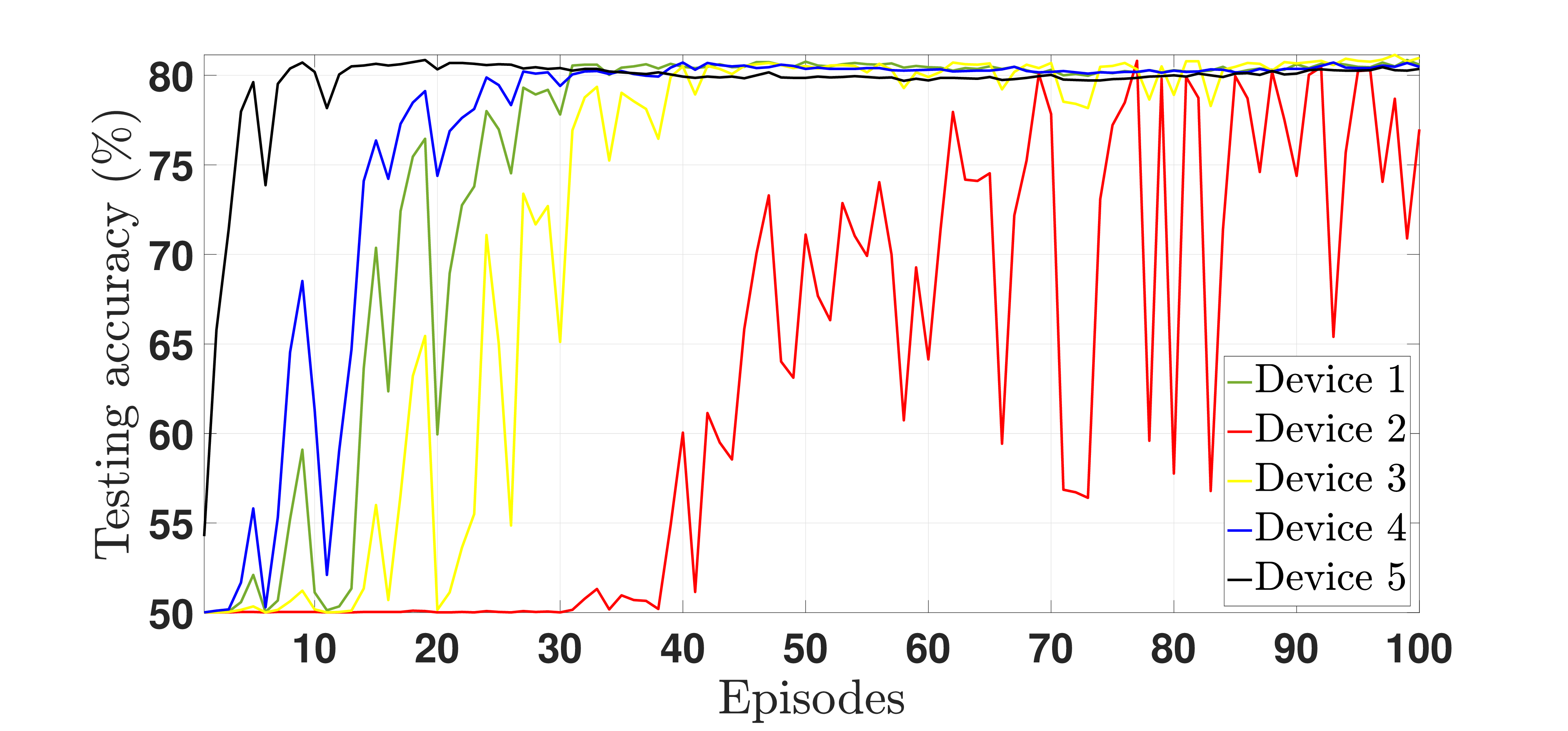} & \includegraphics[width=2.2in]{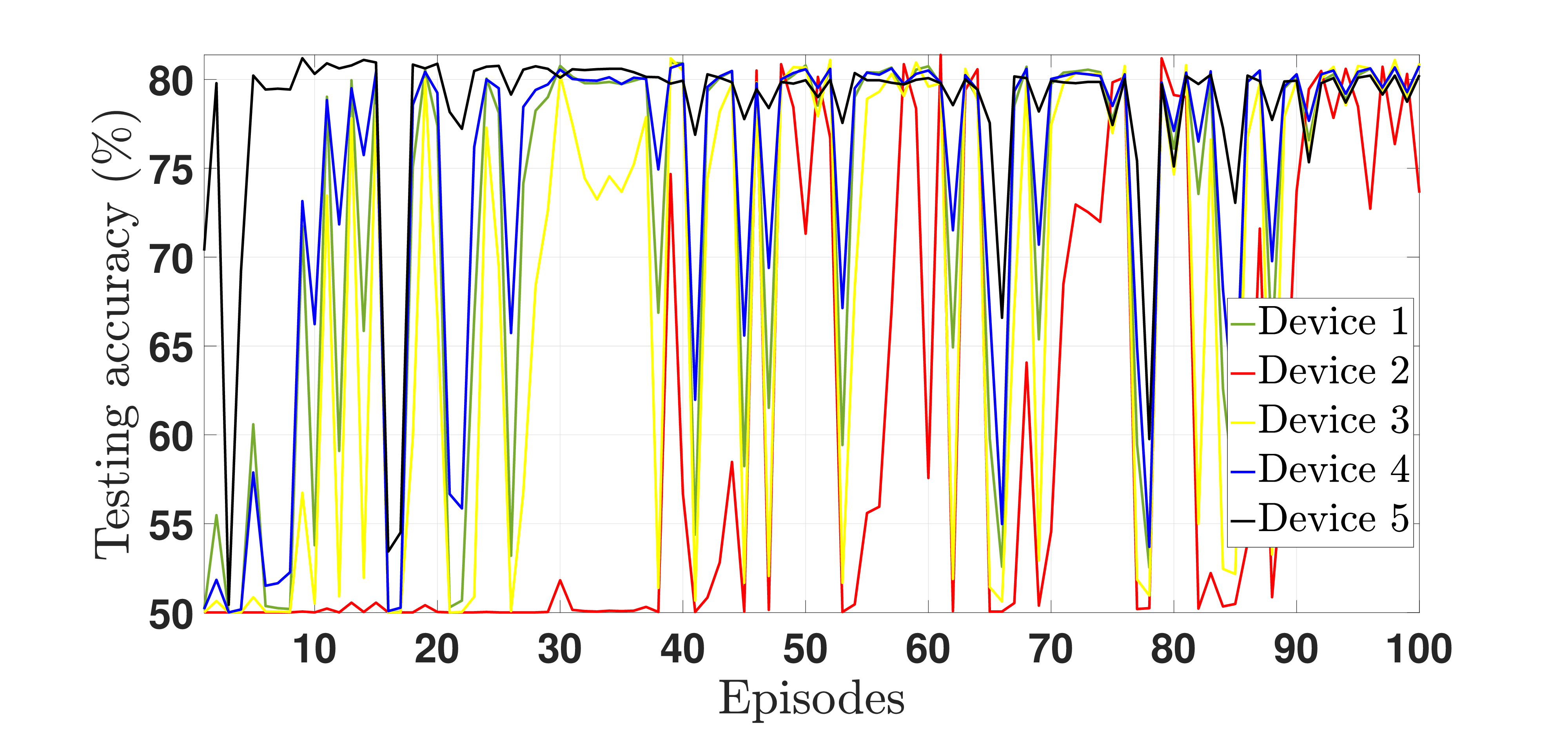} 
\\ (d) $M$ = 100 with FashionMNIST. & (e) $M$ = 200 with FashionMNIST. & (f) $M$ = 300 with FashionMNIST.
\\ \includegraphics[width=2.2in]{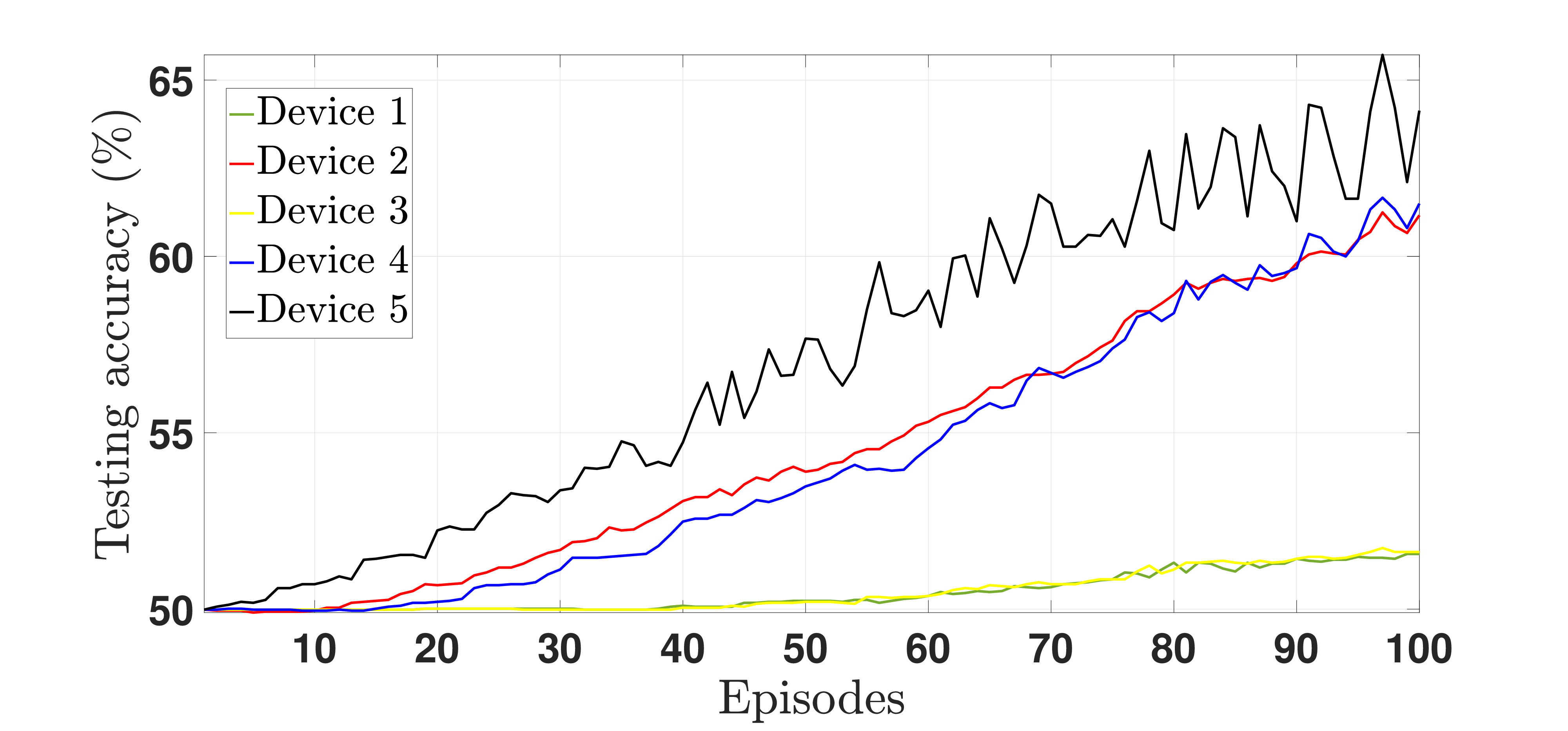} & \includegraphics[width=2.2in]{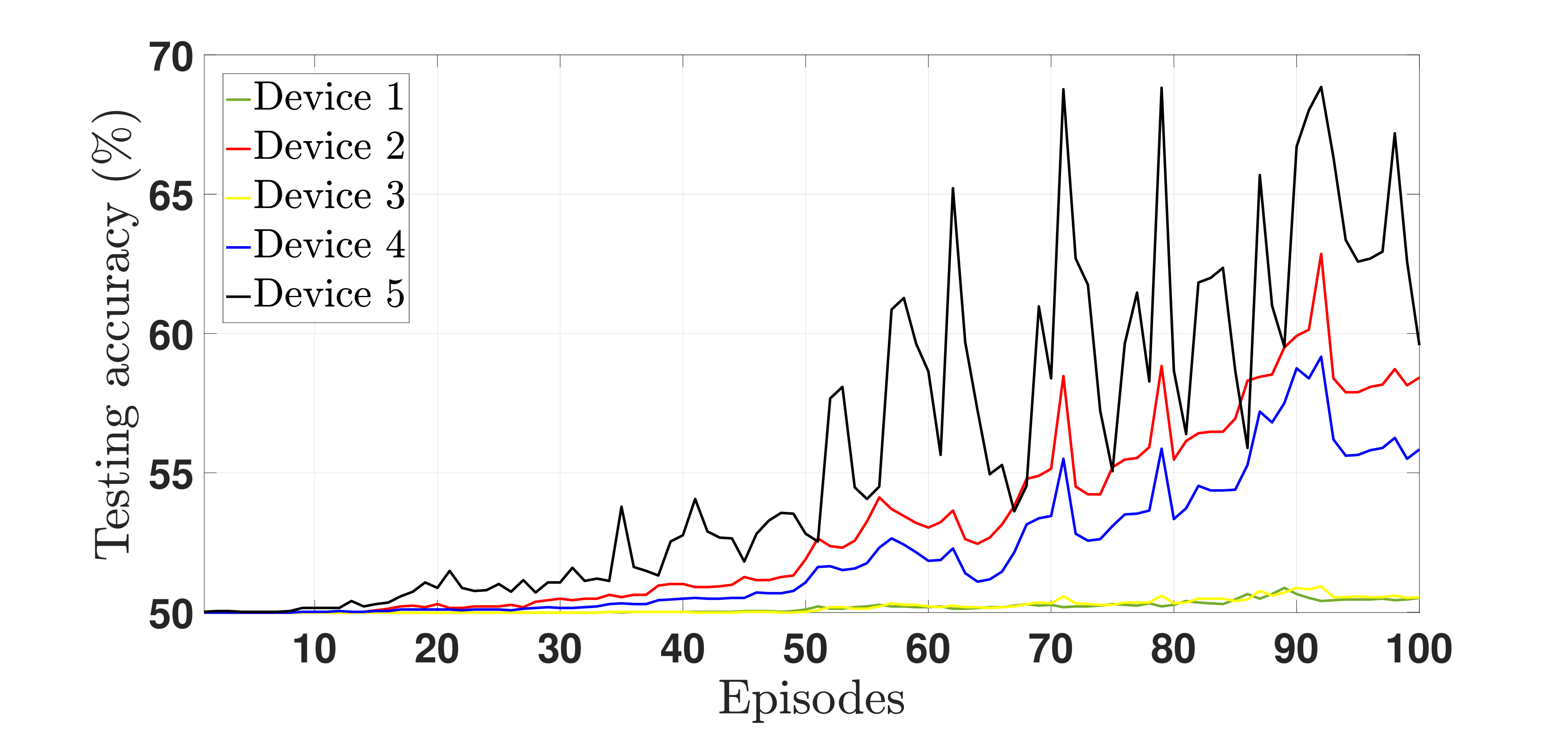} & \includegraphics[width=2.2in]{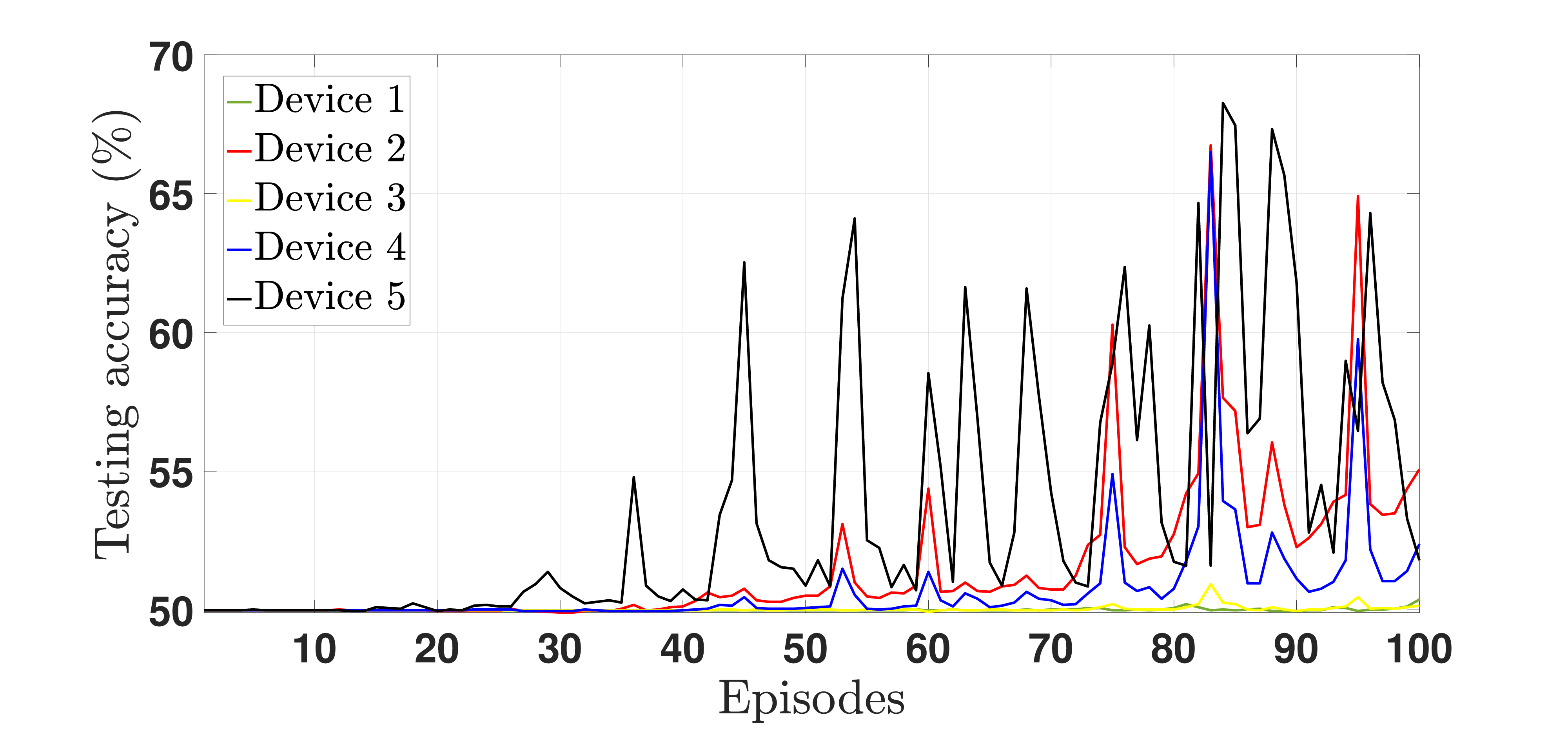} 
\\ (g) $M$ = 100 with with CIFAR-10. & (h) $M$ = 200 with CIFAR-10. & (i) $M$ = 300 with CIFAR-10.
\end{tabular}
\end{center}
\caption{Given 100 FL communication rounds, $I$ = 5 and $J$ = 2, we study the local model's testing accuracy under the proposed VGAE-MP attack on the MNIST, FashionMNIST, and CIFAR-10 datasets.}
\label{fig_accRounds}
\end{figure*}
The benign FL is designed to improve image classification accuracy, while the proposed VGAE-MP attack aims to reduce accuracy and cause label misclassification. 
The number of benign devices $I$ increases from 5 to 30, while the number of attackers $J$ increases from 1 to 5. The global model $\pmb{w}_G^\prime(t)$ in FL is trained with 100 communication rounds, and training of the local model $\pmb{w}_i(t)$ is carried out in 10 iterations. For building the adjacency matrix $\pmb{\mathcal{A}}$ at the attacker, the number of selected model parameters in $\pmb{w}_i(t)$, i.e., $M$, is set to 100, 200, or 300. The VGAE encoder is a two-layer GCN network with a dropout layer to prevent overfitting. The VGAE decoder is an inner product. The Adam optimizer with a learning rate of 0.01 is adopted to optimize the network. For all datasets, we use the same encoder, decoder and SVM models. Table~\ref{sim_config} lists the setting of parameters in PyTorch.

\begin{table} [t]
\centering
\caption{The setting of parameters in PyTorch}
\begin{tabular}{l|l} \hline
\textbf{Parameters} & \textbf{Values}\\ \hline
number of benign devices ($I$) & 5 $\sim$ 30 \\
number of attackers ($J$) & 1 $\sim$ 5 \\
communication rounds of the FL & 100 \\
number of local iterations ($T_L$) & 10 \\
model parameters in $\pmb{w}_i(t)$ ($M$) & 100, 200, or 300 \\
1st hidden layer size of the VGAE & 32 \\
2nd hidden layer size of the VGAE & 16 \\
learning rate of the VGAE & 0.01 \\
batch size of the SVM & 30 \\
learning rate of the SVM & 0.001 \\
regularization of the SVM loss function & 0.01 \\
k-Fold cross-validation & 5 \\ \hline
\end{tabular}
\label{sim_config}
\end{table}

The proposed VGAE-MP attack is implemented on an SVM model using PyTorch 1.12.1, Python 3.9.12 on a Linux workstation with an Intel(R) Core(TM) i7-9700K CPU@3.60GHz (8 cores) and 16 GB of DDR4 memory@2400 MHz. 
The experiments are carried out on three datasets:
\begin{itemize}
\item The standard MNIST dataset, comprising 60,000 training and 10,000 testing  grayscale images of handwritten digits from 1 to 10;
\item The FashionMNIST dataset, comprising Zalando's article grayscale images with the size of 28 $\times$ 28 in ten classes, including 60,000 and 10,000 images for training and testing, respectively;
\item The CIFAR-10 dataset, consisting of 60,000 images with the size of 32 $\times$ 32 in ten classes (6,000 per class), 50,000 for training and 10,000 for testing.
\end{itemize}

At each benign user device, a standard quadratic optimization algorithm is utilized to train the SVM models with the three datasets. The loss function of the SVM models is $F_i(\pmb{w}_i(t)) = \frac{1}{2}\left\|\pmb{w}_i(t) \right\|_2^2 + \frac{1}{D_i}\sum_{i=1}^{I}\max\left\lbrace 0, 1 - y_i^{d_i} ({\beta}_i+ {\pmb{\omega}^T_i(t)} x_i^{d_i}) \right\rbrace$, where ${\beta}_i$ is a parameter that can be obtained based on ${\pmb{w}_i(t)}$.

\begin{figure*}[!t]
\begin{center}
\begin{tabular}{ccc}
\includegraphics[width=2.2in]{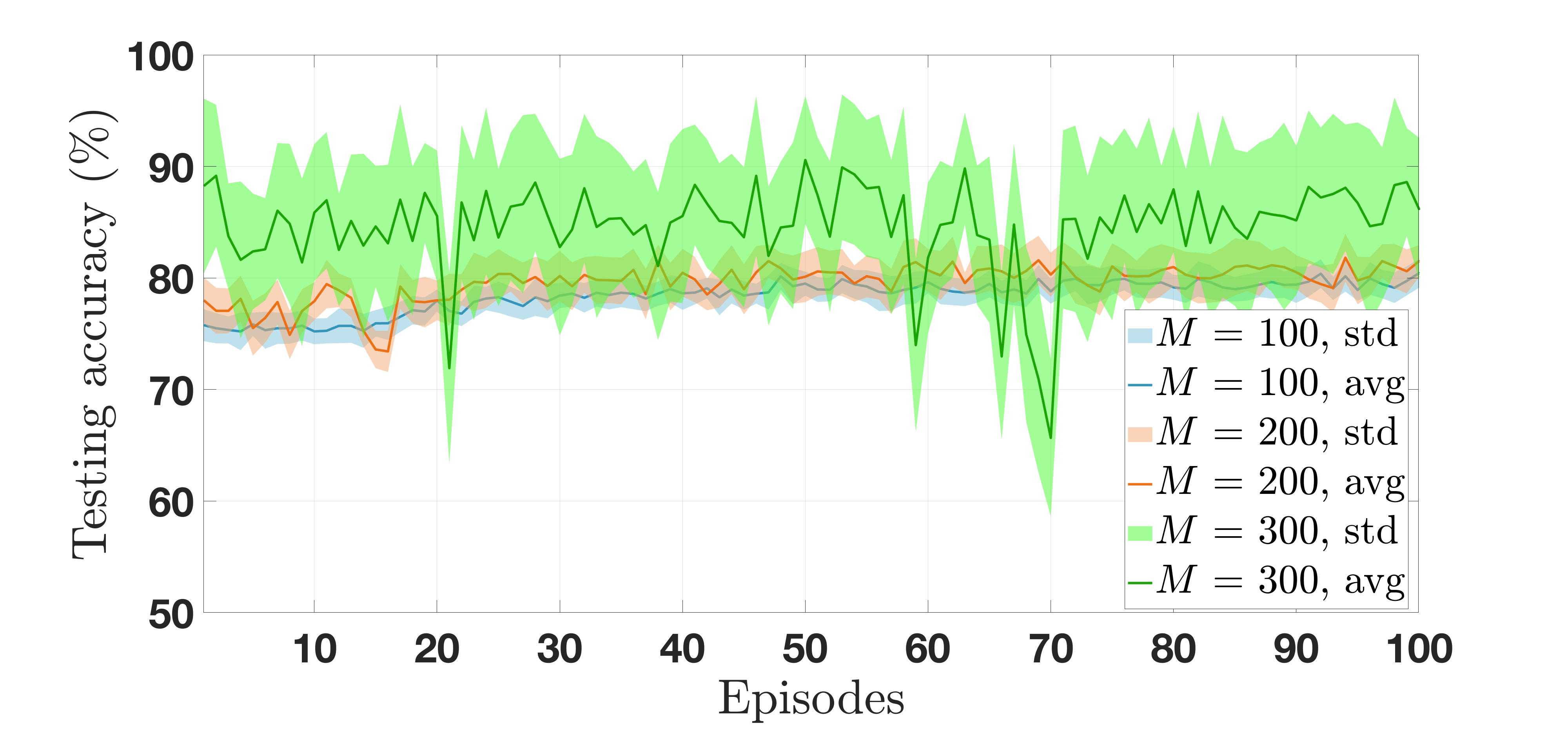} & \includegraphics[width=2.2in]{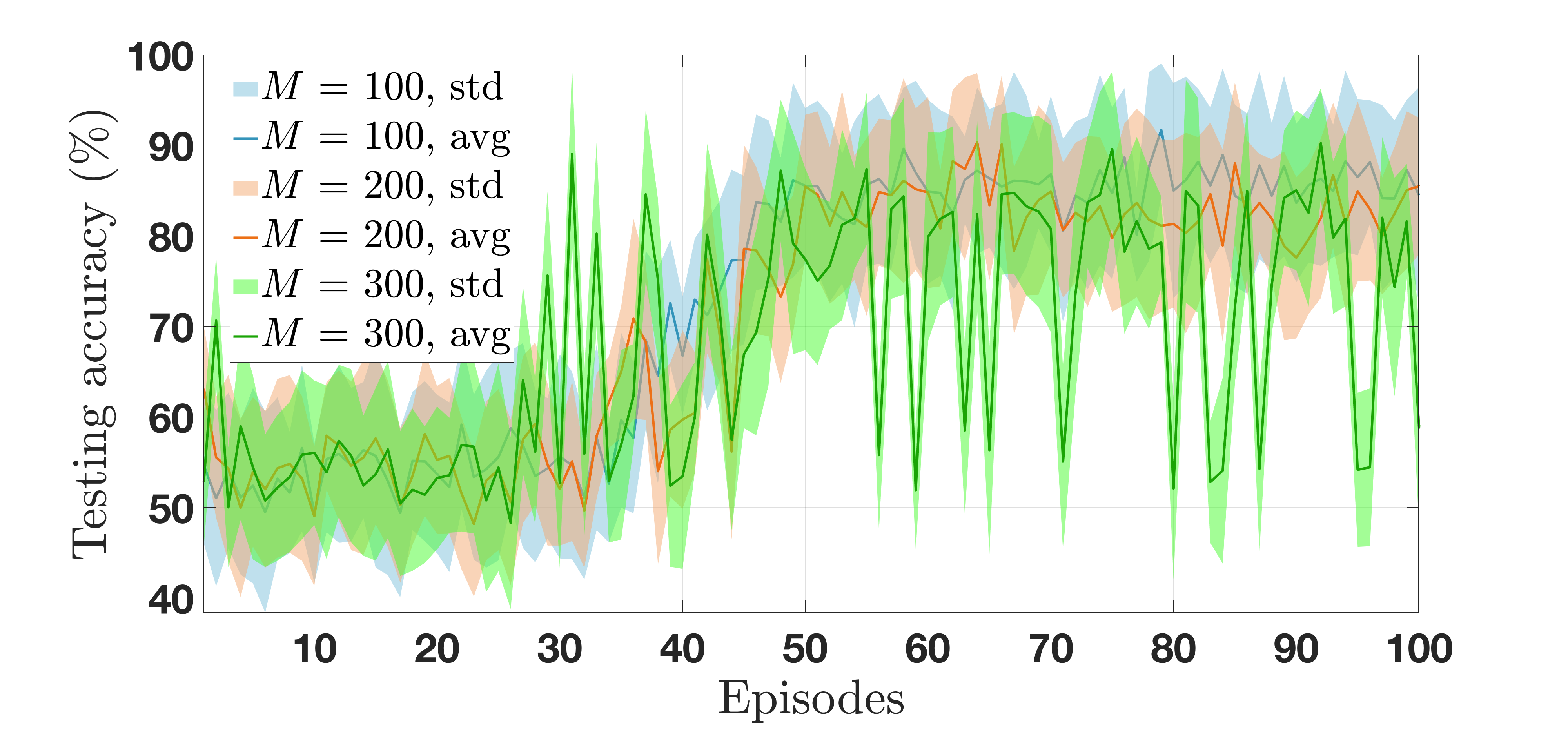} & \includegraphics[width=2.2in]{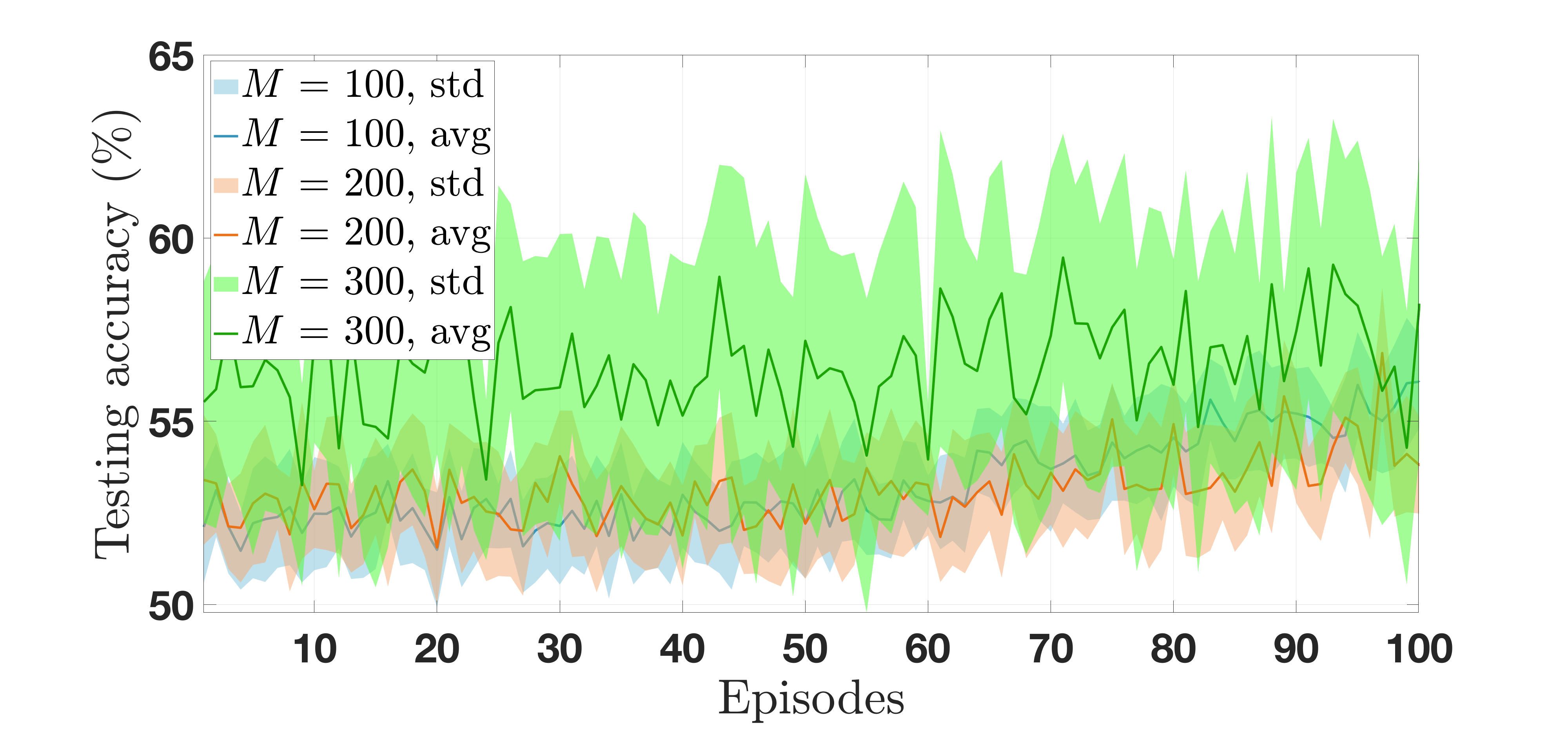}\\
(a) MNIST & (b) FashionMNIST & (c) CIFAR-10 \\
\end{tabular}
\end{center}
\caption{The global model's testing accuracy (``avg'' means the average value and ``std'' stands for the standard deviation) under the VGAE-MP attack on the MNIST, FashionMNIST, and CIFAR-10 datasets.}
\label{fig_accGlobal}
\end{figure*}

\begin{figure*}[!t]
\begin{center}
\begin{tabular}{ccc}
\includegraphics[width=2.2in]{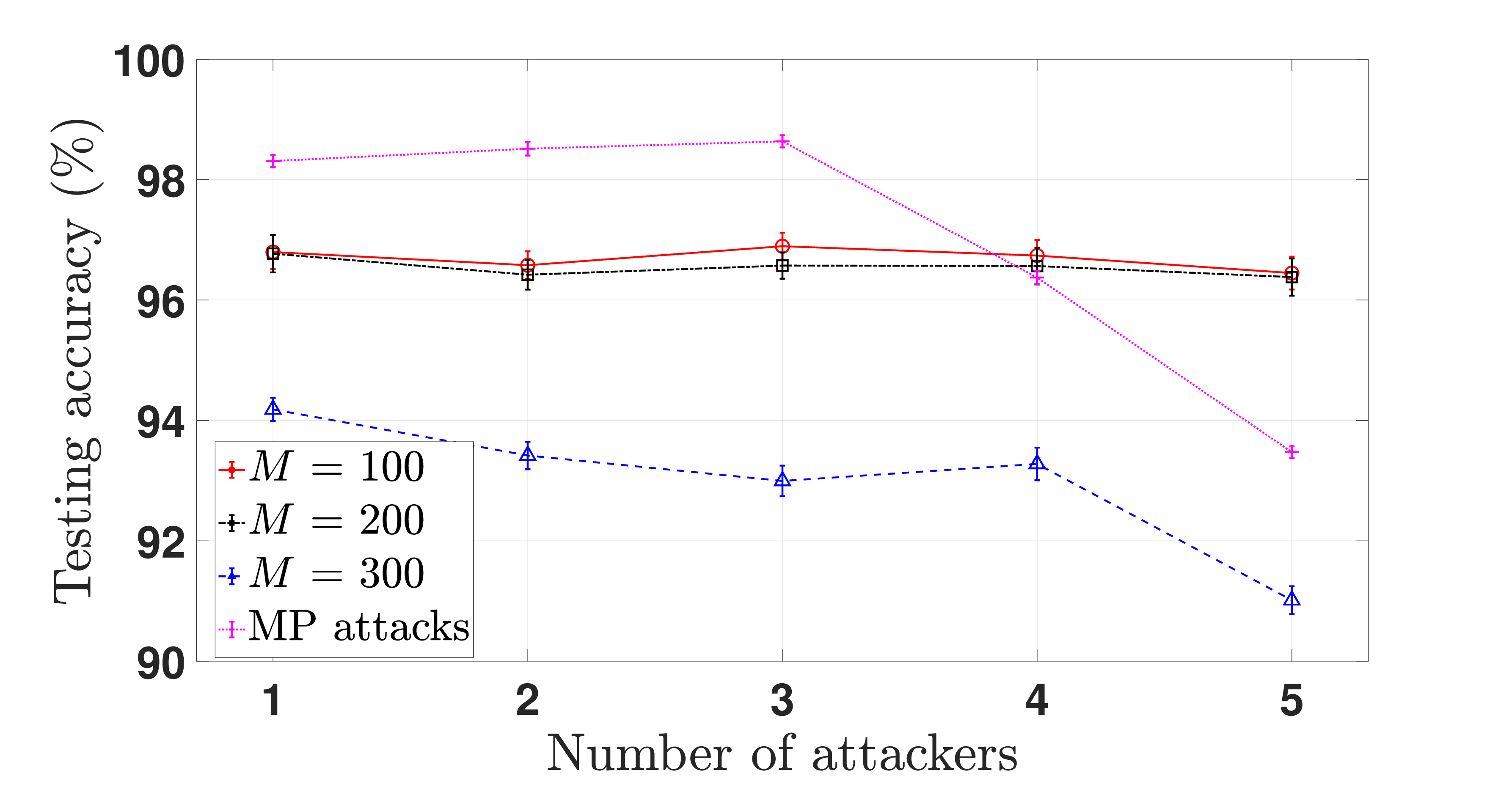} & \includegraphics[width=2.2in]{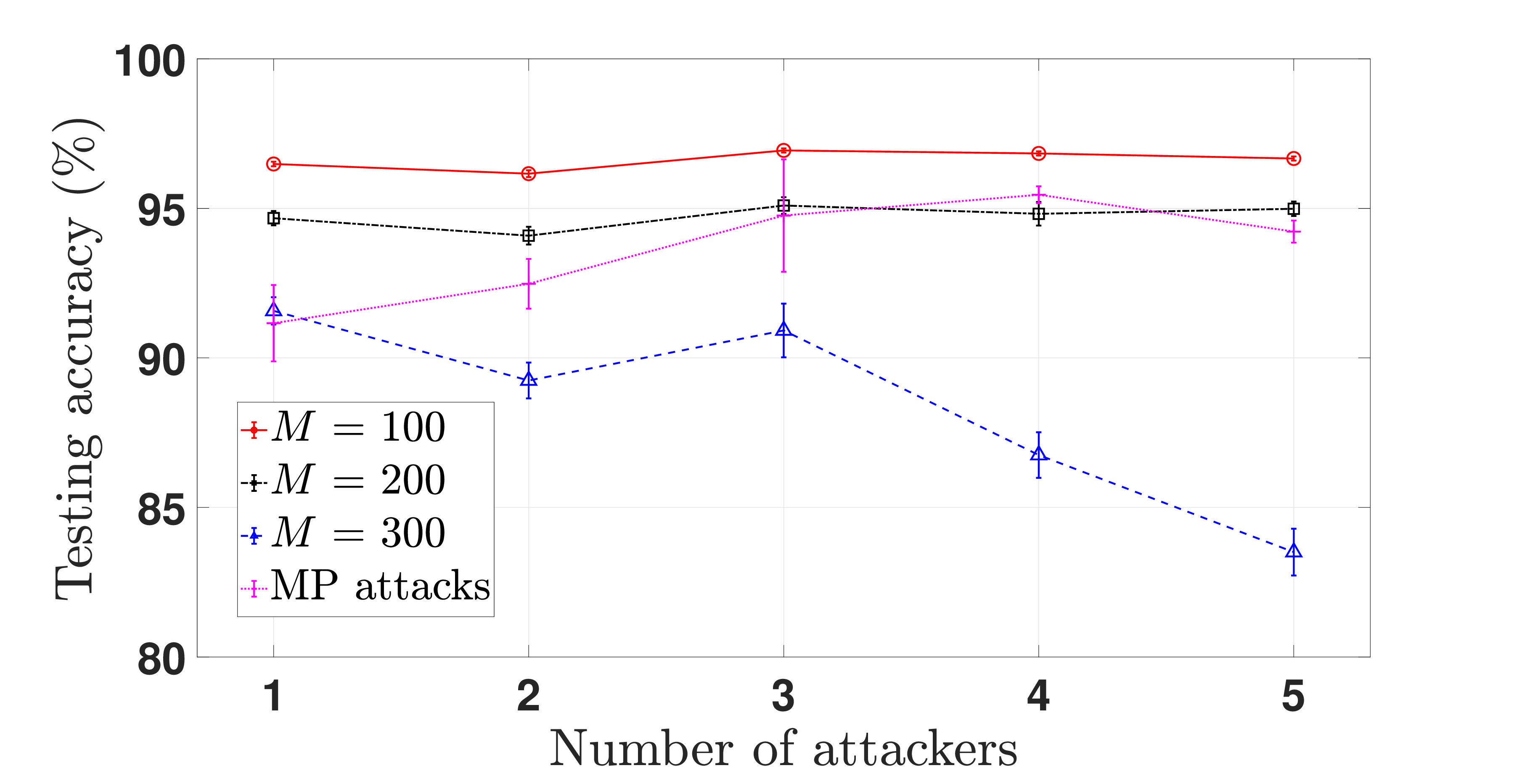} & \includegraphics[width=2.2in]{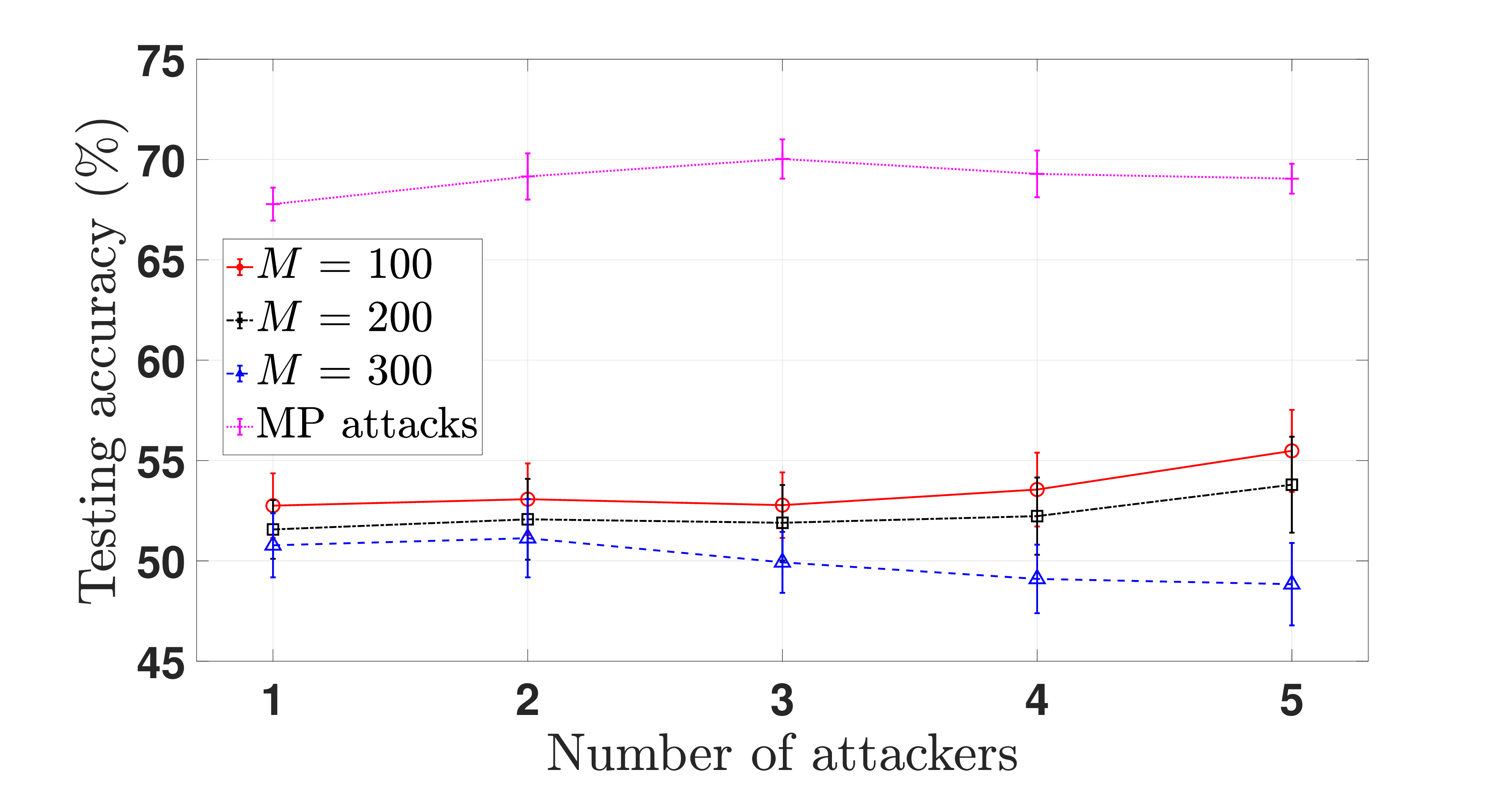}\\
(a) MNIST & (b) FashionMNIST & (c) CIFAR-10 \\
\end{tabular}
\end{center}
\caption{Given the MNIST, FashionMNIST, and CIFAR-10 datasets, the average testing accuracy of the local models under the VGAE-MP attack when $J$ increases from 1 to 5.}
\label{fig_attackers}
\end{figure*}

\begin{figure*}[!t]
\begin{center}
\begin{tabular}{ccc}
\includegraphics[width=2.2in]{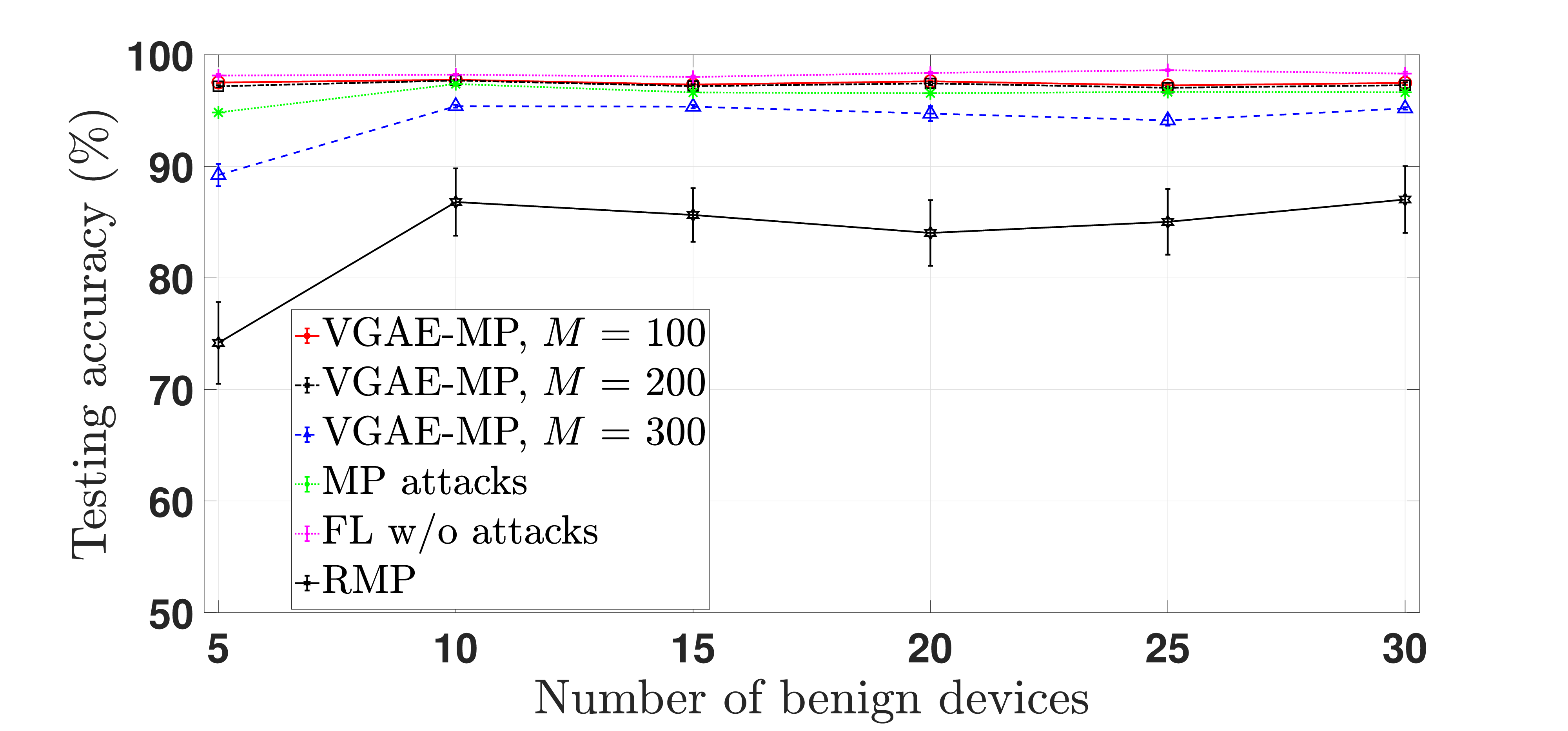} & \includegraphics[width=2.2in]{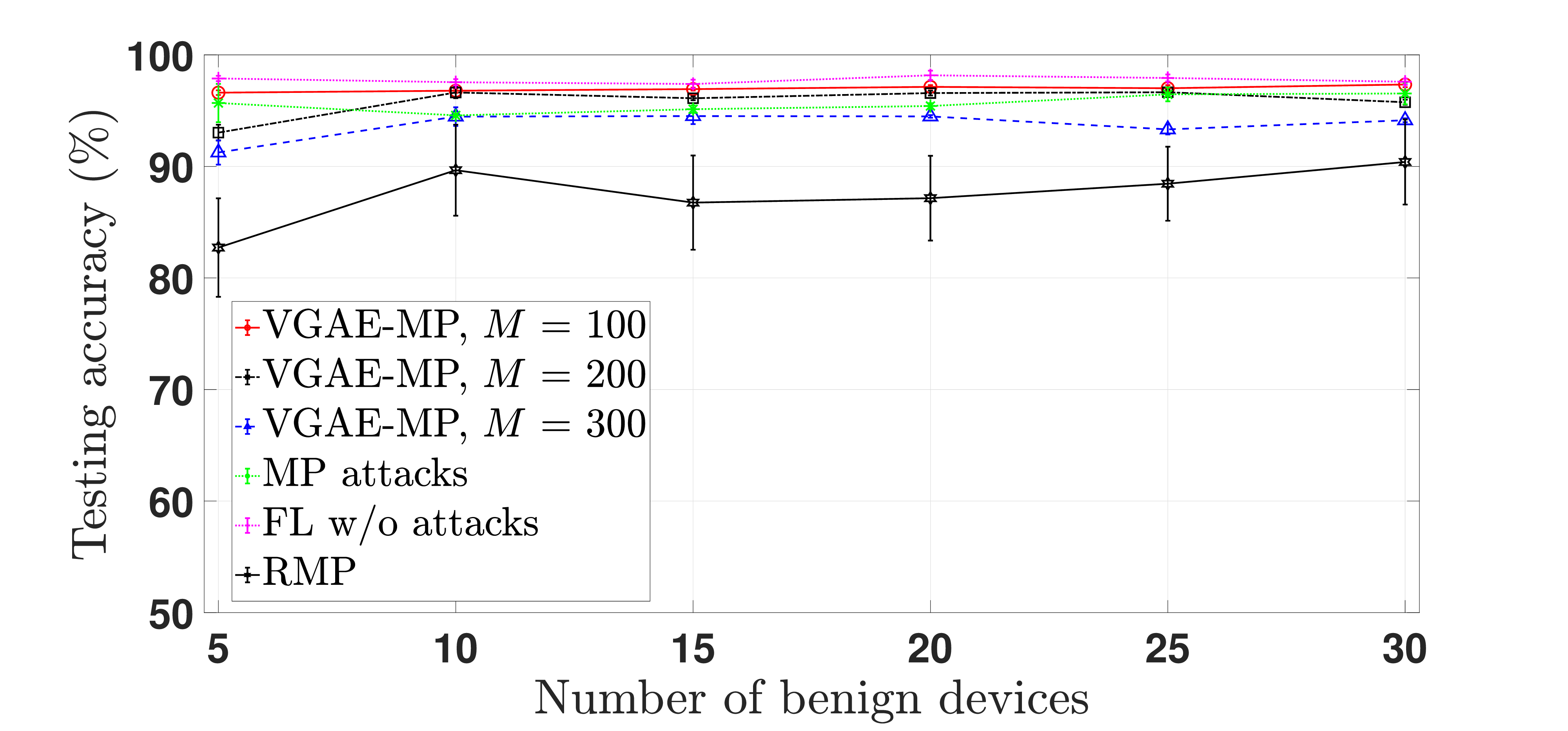} & \includegraphics[width=2.2in]{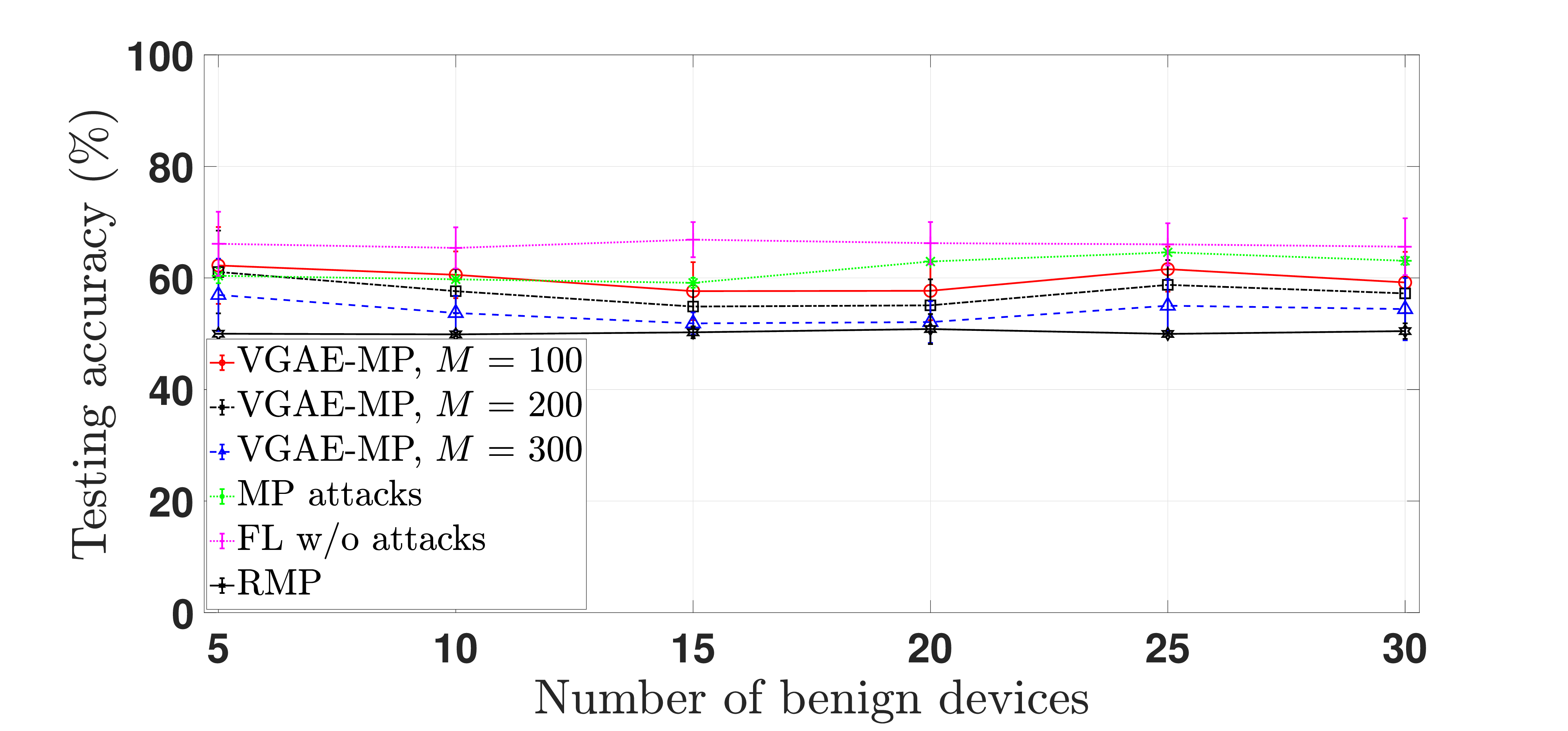}\\
(a) MNIST & (b) FashionMNIST & (c) CIFAR-10 \\
\end{tabular}
\end{center}
\caption{Given $J$ = 5, the average testing accuracy under the VGAE-MP attack on the MNIST, FashionMNIST, and CIFAR-10 datasets, where $I$ increases from 5 to 30.}
\label{fig_accDevices}
\end{figure*}

In addition, the proposed VGAE-MP attack is compared with an existent data-agnostic model poisoning (MP) attack that produces malicious local models by mimicking other benign devices’ training samples to degrade the learning accuracy. The MP attack considered for comparison has been used in several existing studies, e.g.,~\cite{zheng2024detecting,hossain2021desmp,cao2022flcert}, where the attacker manipulates the training process by injecting a fake device and sending fake local models to the server. Moreover, we implemented another existing attack on FL, i.e., a random MP (RMP) attack considered in~\cite{fang2020local,cao2022mpaf,chen2024exploring}. Specifically, RMP generates the malicious local model by injecting a Gaussian random noise into the received global model, which can enlarge the magnitudes of the random local model updates using a scaling factor. 

\subsection{Attacking Performance}
In Fig.~\ref{fig_accRounds}, we plot the local model's testing accuracy with 100 FL communication rounds under the proposed VGAE-MP attack on the MNIST, FashionMNIST, and CIFAR-10. When $M$ of VGAE-MP increases from 100 to 300, the FL accuracy fluctuates dramatically, successfully restraining the convergence of the testing accuracy. Using FashionMNIST as an example, the FL accuracy of the five benign devices converges to 80\% in~\ref{fig_accRounds}(d) under the VGAE-MP attack with $M$ = 100. Once $M$ increases to 300 in~\ref{fig_accRounds}(f), the FL accuracy of the five benign devices consistently experiences fluctuations between 50\% and 80\%. This confirms that $M$ determines the size of features in $\pmb{\omega}^m(t)$ whose correlation in $\pmb{\mathcal{A}}$ is learned to generate the malicious poisoning model $\pmb{w}_j^\prime(t)$. Therefore, a large $M$ leads to a more complete graph trained by the VGAE model. 

In Figs.~\ref{fig_accRounds}(a) to~\ref{fig_accRounds}(i), we interestingly observe that VGAE-MP demonstrates more prominent attacking performance on the FL with the FashionMNIST and CIFAR-10 than the one with the MNIST. This might be attributed to the variances in the MNIST, FashionMNIST, and CIFAR-10. MNIST comprises grayscale images of handwritten digits, whereas FashionMNIST houses grayscale images of apparel and accessories. CIFAR-10, on the other hand, has 10 distinct categories of objects, including animals, vehicles, among others. The simplicity of the MNIST's handwritten digits may make them more easily classified by FL compared to the more complex images found in FashionMNIST or CIFAR-10.

Fig.~\ref{fig_accGlobal} shows the global model's testing accuracy measured at the server based on the MNIST, FashionMNIST, and CIFAR-10. Under the VGAE-MP attack, the steady convergence of FL accuracy is inhibited. In particular, for the CIFAR-10  with $M$ = 300, the FL accuracy maintains around 58\% under the VGAE-MP attack. Moreover, the VGAE-MP attack doesn't lead to a considerable decrease in the testing accuracy of the global model. This is because that a significant performance dip could potentially reveal the presence of the attacker.

\begin{figure*}[htb]
\begin{center}
\begin{tabular}{cc}
\includegraphics[width=3.5in]{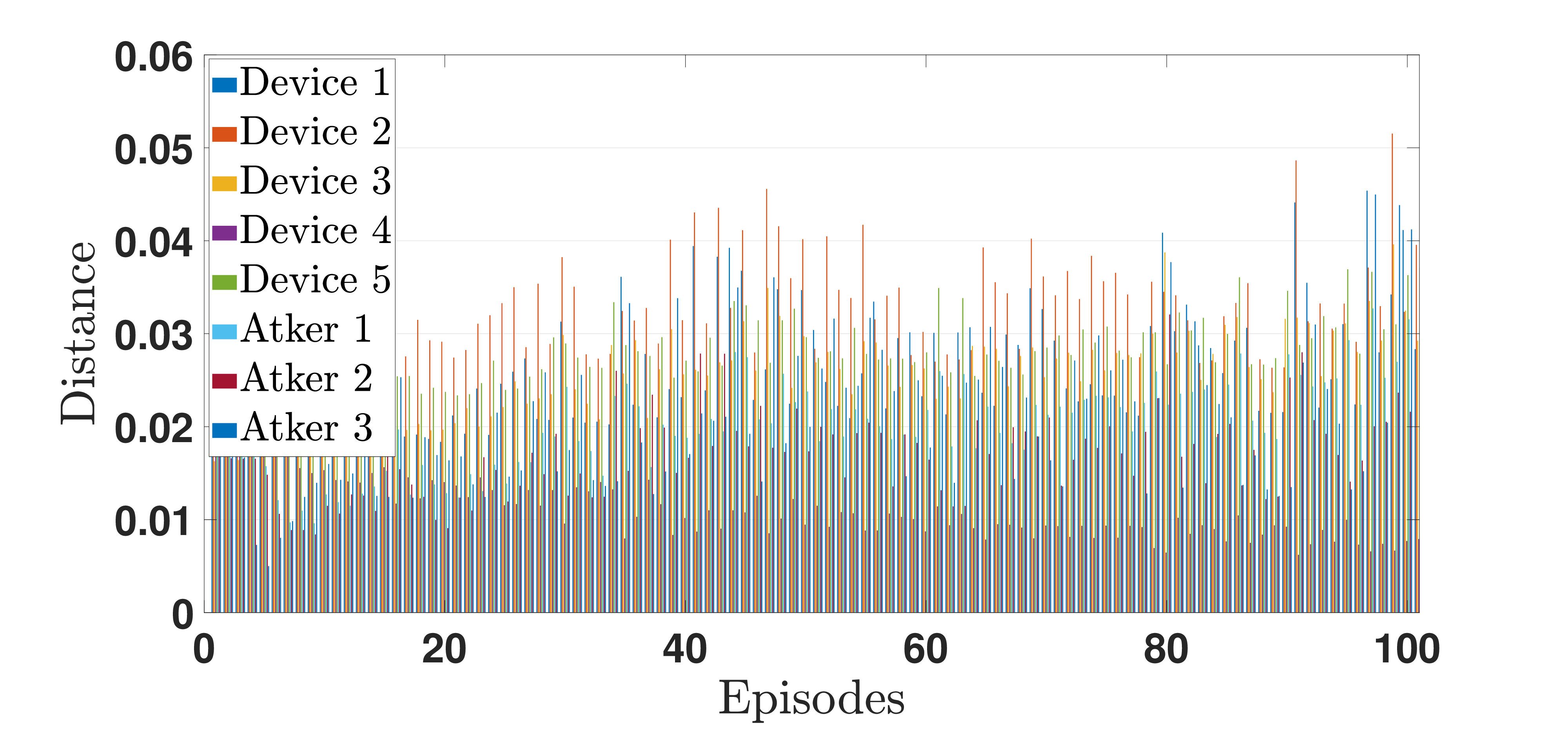} & \includegraphics[width=3.5in]{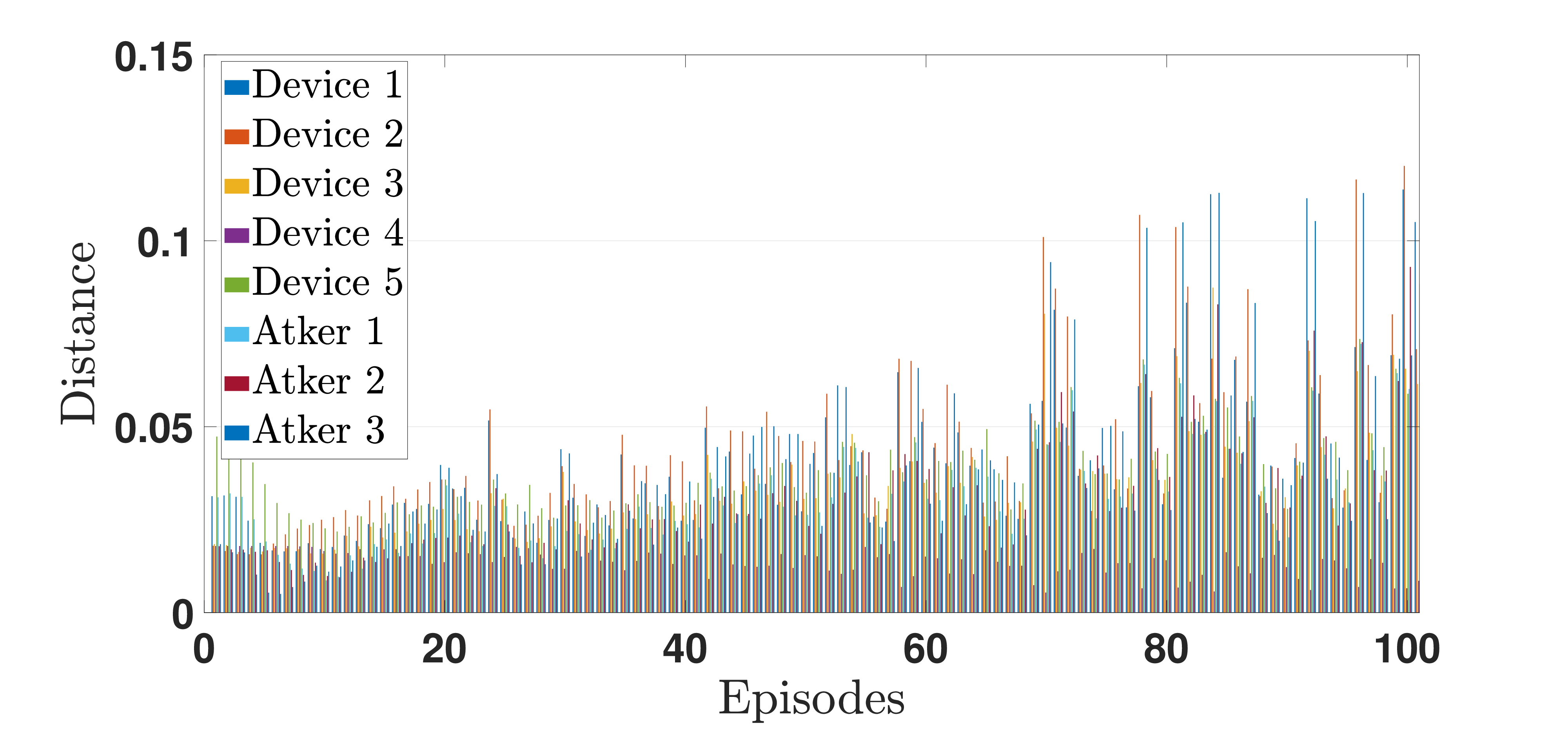} \\
(a) VGAE-MP with $M$ = 100 & (b) VGAE-MP with $M$ = 200 \\
\includegraphics[width=3.5in]{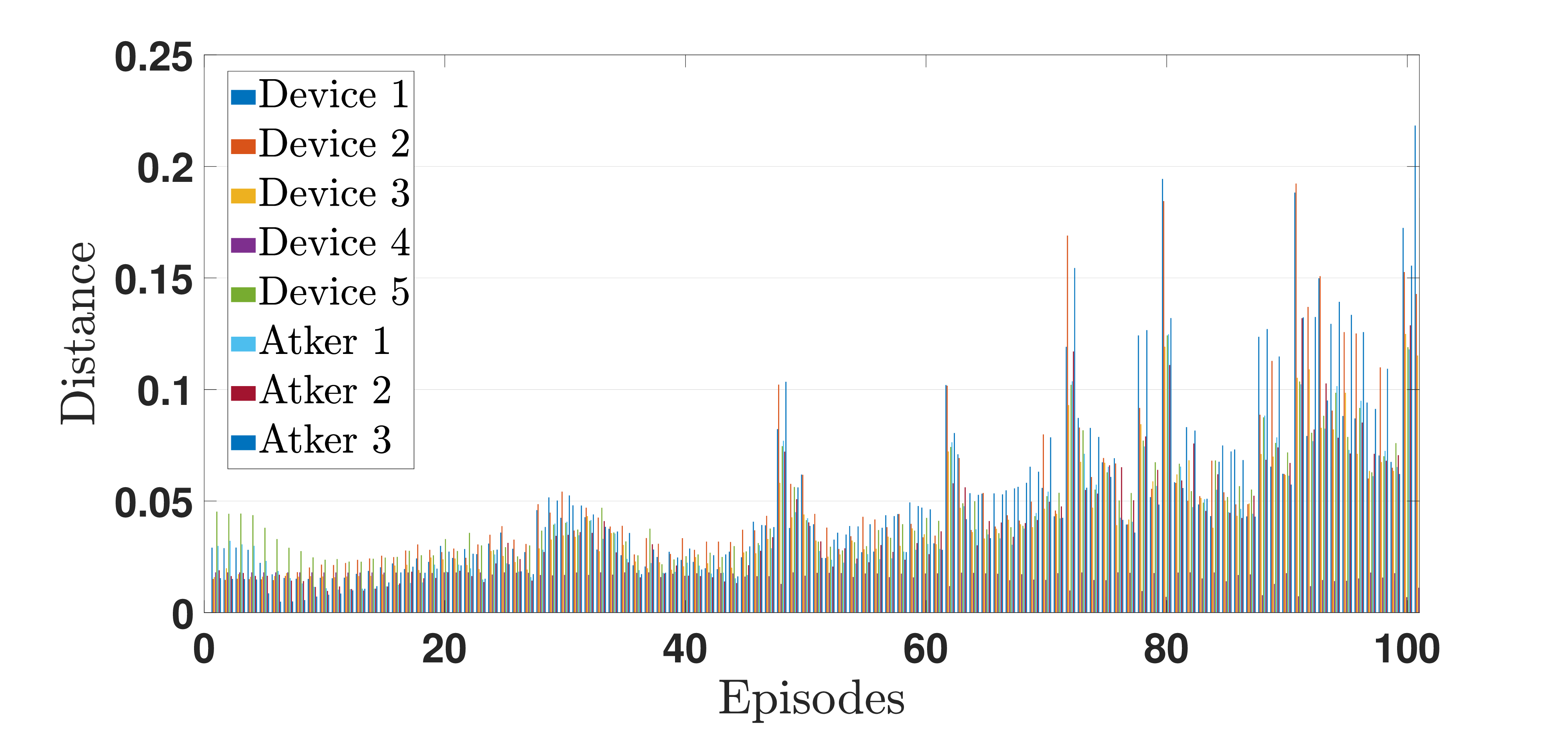} & \includegraphics[width=3.5in]{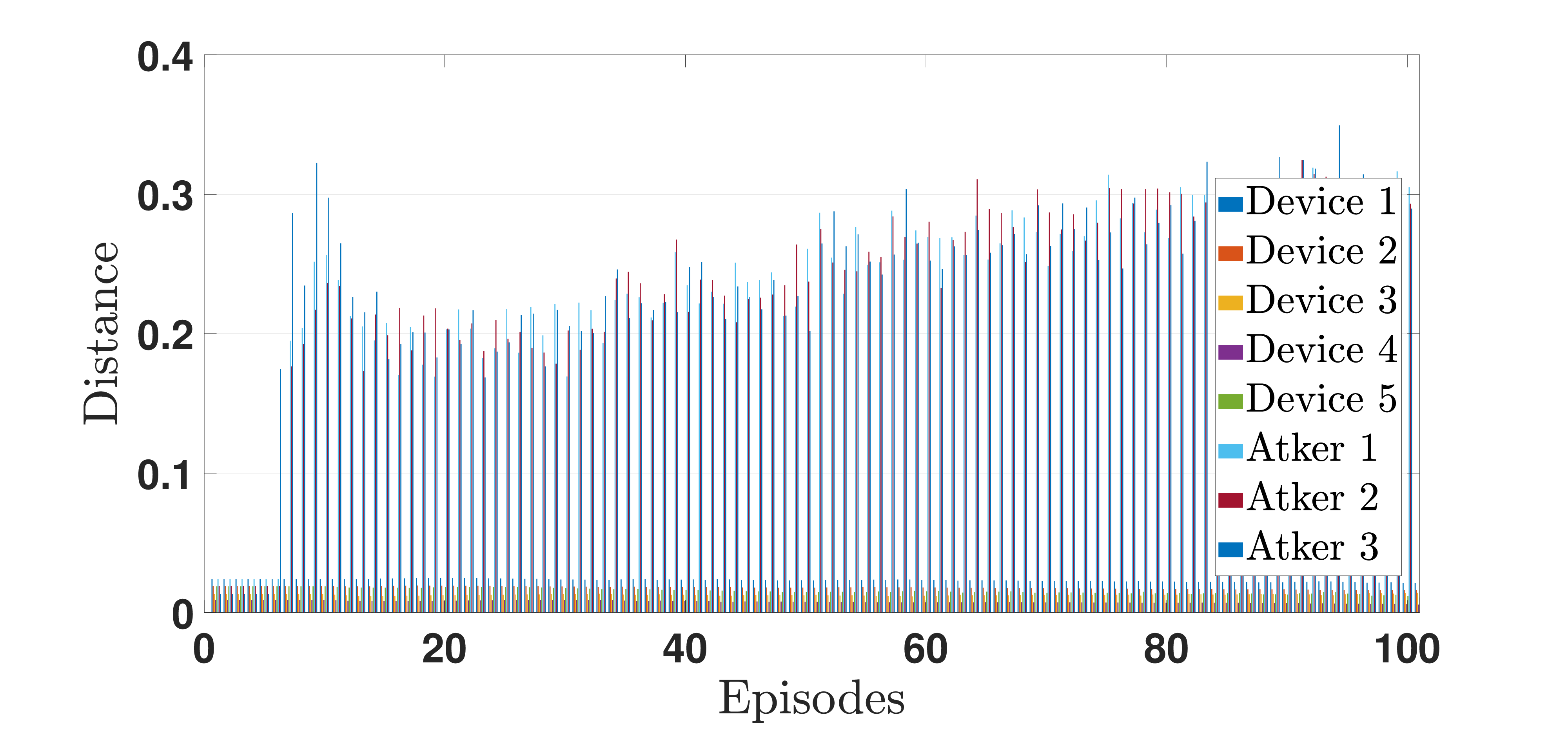} \\
(c) VGAE-MP with $M$ = 300 & (d) The existing MP attack~\cite{hossain2021desmp} 
\end{tabular}
\\ \includegraphics[width=3.5in]{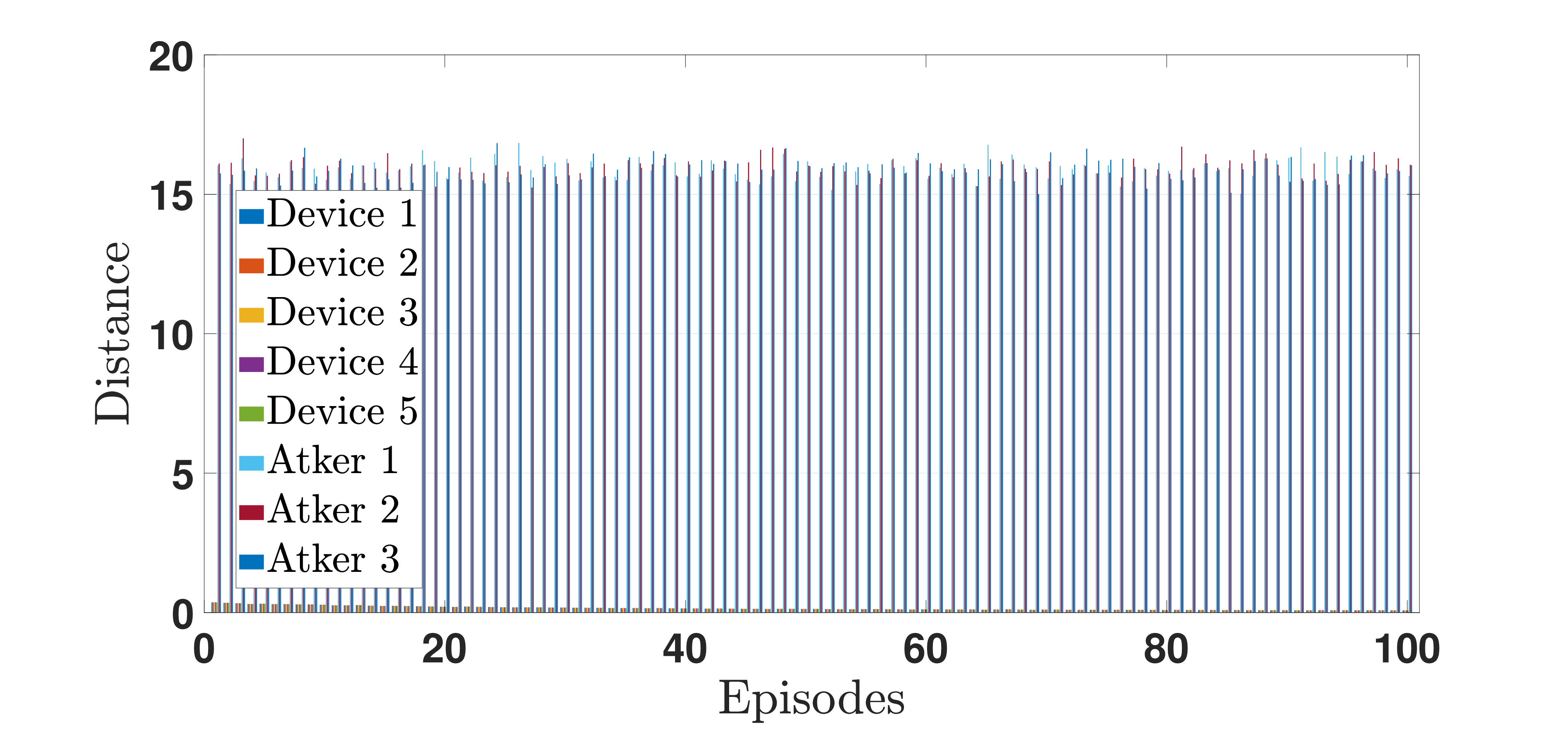} \\
(e) The RMP attack~\cite{cao2022mpaf}
\end{center}
\caption{Based on the CIFAR-10 training datasets, the Euclidean distances of the local models are measured at the server in order to detect a poisoning attack, where we set $I$ = 5 and $J$ = 3.}
\label{fig_distance}
\end{figure*}

Fig.~\ref{fig_attackers} plots the average testing accuracy of the local models under the VGAE-MP attack when $J$ increases from 1 to 5. Since the number of benign devices is fixed at 3, the FL accuracy falls with the growth of the number of attackers. This is because the proposed VGAE-MP attack hinders the training convergence of FL. In particular, when $M$ = 300, the average testing accuracy under the VGAE-MP attack drops about 5\%, 12\%, and 4\% according to the MNIST, FashionMNIST, and CIFAR-10, respectively. When $J$ = 5, the VGAE-MP attack outperforms the MP attack 10\% and 20\% given the FashionMNIST and CIFAR-10, respectively. The reason is the new VGAE-MP attack reconstructs the adversarial adjacency matrix according to the individual features of the devices. Consequently, the attacker falsifies the local models to maximize the FL loss. 

Fig.~\ref{fig_accDevices} depicts the average testing accuracy of FL without the attack and FL under the VGAE-MP, MP or RMP attack, where $J$ is set to 5 and $I$ ranges from 5 to 30. As the benign devices increase, the FL accuracy under the VGAE-MP, MP and RMP attacks improves, given that the FL can quickly converge when the ratio of $\pmb{w}_i(t)$ to $\pmb{w}_j^\prime(t)$ is heightened. On the three considered datasets, the FL accuracy is 6\%, 5\%, and 5\% under the VGAE-MP attack lower than it is under the MP attack, respectively, when $M$ = 300 and $I$ = 5. The RMP attack has lower FL accuracy than the VGAE-MP and MP attacks. This is because the malicious local model update of the RMP attack is generated according to a Gaussian random noise, which is not correlated with any benign local models. However, this can make the malicious local models more easily detected and subsequently eliminated, as will be shown in Fig.~\ref{fig_distance}.

Existing MP attacks in FL result in a high training loss of the FL model. One way to detect these malicious attacks is to compare the distance between the malicious local and global models with the distance between the benign local and global models. Suppose the distance between the malicious local and global models is larger. In such case, it can indicate a malicious attack, and the server can detect it accordingly. 

To evaluate the invisibility of the proposed VGAE-MP attack, we study the distance between the local and the global models based on the CIFAR-10 datasets in Fig.~\ref{fig_distance}, where $I$ = 5 and $J$ = 3. As shown in Figs.~\ref{fig_distance}(a),~\ref{fig_distance}(b), and~\ref{fig_distance}(c), the Euclidean distances between the malicious local models generated by the new VGAE-MP attack and the corresponding global models are below that of the benign local models. This makes it difficult for the server to detect and defend against the attacker. In contrast, as shown in Figs.~\ref{fig_distance}(d) and~\ref{fig_distance}(e), the MP attack and the RMP attack result in a significantly larger distance between the malicious local and global models, making them easier to detect. This highlights the key strength of the proposed VGAE-MP attack, that is, VGAE-MP generates malicious local models based on the feature correlation between the benign local and global models, and hence makes the differences between the malicious and benign local models indistinguishable. 

\begin{figure}[htb]
\centering
\includegraphics[width=3.6in]{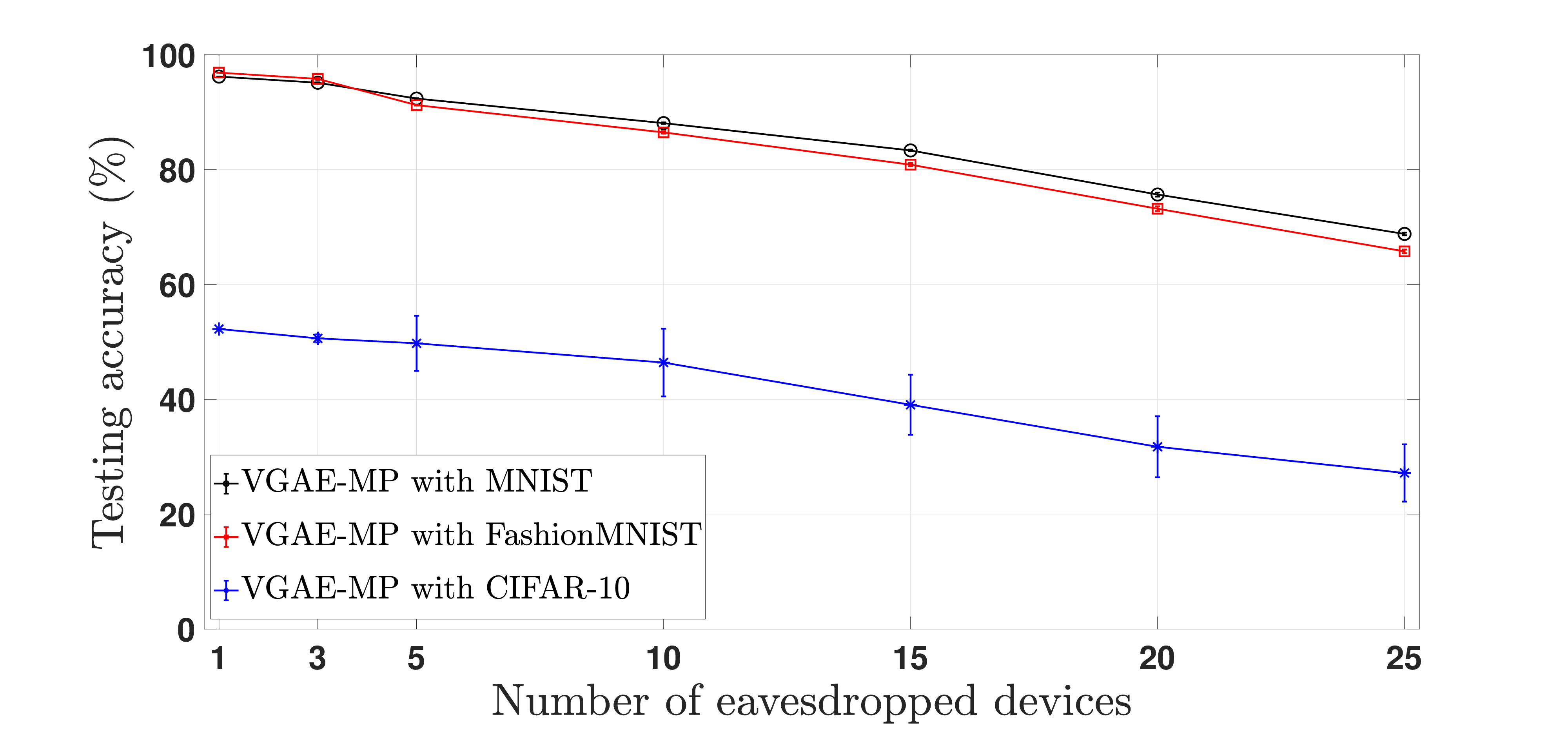}
\caption{The number of eavesdropped benign local model updates increases from 1 to 25, based on the MNIST, FashionMNIST, or CIFAR-10 datasets.}
\label{fig_accOverheard}
\end{figure}

Fig.~\ref{fig_accOverheard} shows the average testing accuracy under the proposed VGAE-MP attack on the MNIST, FashionMNIST, or CIFAR-10 dataset. This is observed as the attacker eavesdrops on an increasing number of benign user devices, ranging from 1 to 25. Generally, a noticeable fall in the local model updates' average accuracy is observed as the number of eavesdropped benign devices escalates. This trend is attributed to the attacker's ability to intercept more benign local models, thereby acquiring a broader range of correlation features. Such extensive data aids in crafting a more potent malicious model for effective system poisoning. In particular, the average accuracy on the MNIST, FashionMNIST, and CIFAR-10 datasets drops about 27.4\%, 32.3\%, and 24.9\%, respectively.

\section{Conclusion and Future Work}
\label{sec_cond}
In this paper, a new data-untethered VGAE-MP attack against FL was proposed, where the adversarial VGAE was developed to create malicious local models based solely on the benign local models overheard without access to the training data of FL. The proposed adversarial VGAE allows the attacker to extract the common underlying data features of the benign local models and their correlations to generate the malicious model with which the FL training loss is maximized. The VGAE-MP attack maintains the feature correlation between the benign local and global models, making the differences between the malicious and benign local models indistinguishable. The VGAE-MP attack on the FL was implemented using PyTorch with the source code released on GitHub. The performances were evaluated using the MNIST, fashionMNIST, and CIFAR-10 datasets. 

The proposed data-untethered VGAE-MP attack involves a single poisoning objective, which aims to degrade the training accuracy of FL. In our future work, multiple performance metrics of FL will be considered in the poisoning, such as training fairness, robustness, and model utility. A multi-objective optimization will be formulated while the VGAE will be further studied to extract the graph representation.

\ifCLASSOPTIONcaptionsoff
  \newpage
\fi

\bibliographystyle{IEEEtran}
\bibliography{bibVGAE}  

\begin{thebibliography}{10}
\providecommand{\url}[1]{#1}
\csname url@samestyle\endcsname
\providecommand{\newblock}{\relax}
\providecommand{\bibinfo}[2]{#2}
\providecommand{\BIBentrySTDinterwordspacing}{\spaceskip=0pt\relax}
\providecommand{\BIBentryALTinterwordstretchfactor}{4}
\providecommand{\BIBentryALTinterwordspacing}{\spaceskip=\fontdimen2\font plus
\BIBentryALTinterwordstretchfactor\fontdimen3\font minus
  \fontdimen4\font\relax}
\providecommand{\BIBforeignlanguage}[2]{{%
\expandafter\ifx\csname l@#1\endcsname\relax
\typeout{** WARNING: IEEEtran.bst: No hyphenation pattern has been}%
\typeout{** loaded for the language `#1'. Using the pattern for}%
\typeout{** the default language instead.}%
\else
\language=\csname l@#1\endcsname
\fi
#2}}
\providecommand{\BIBdecl}{\relax}
\BIBdecl

\bibitem{tan2022towards}
A.~Z. Tan, H.~Yu, L.~Cui, and Q.~Yang, ``Towards personalized federated
  learning,'' \emph{IEEE Transactions on Neural Networks and Learning Systems},
  2022.

\bibitem{li2022internet}
K.~Li, Y.~Cui, W.~Li, T.~Lv, X.~Yuan, S.~Li, W.~Ni, M.~Simsek, and F.~Dressler,
  ``When internet of things meets metaverse: Convergence of physical and cyber
  worlds,'' \emph{IEEE Internet of Things Journal}, vol.~10, no.~5, pp.
  4148--4173, 2022.

\bibitem{lyu2022privacy}
L.~Lyu, H.~Yu, X.~Ma, C.~Chen, L.~Sun, J.~Zhao, Q.~Yang, and S.~Y. Philip,
  ``Privacy and robustness in federated learning: Attacks and defenses,''
  \emph{IEEE Transactions on Neural Networks and Learning Systems}, 2022.

\bibitem{jagielski2018manipulating}
M.~Jagielski, A.~Oprea, B.~Biggio, C.~Liu, C.~Nita-Rotaru, and B.~Li,
  ``Manipulating machine learning: Poisoning attacks and countermeasures for
  regression learning,'' in \emph{IEEE symposium on security and privacy
  (SP)}.\hskip 1em plus 0.5em minus 0.4em\relax IEEE, 2018, pp. 19--35.

\bibitem{wang2022poisoning}
Z.~Wang, Y.~Huang, M.~Song, L.~Wu, F.~Xue, and K.~Ren, ``Poisoning-assisted
  property inference attack against federated learning,'' \emph{IEEE
  Transactions on Dependable and Secure Computing}, 2022.

\bibitem{gao2021secure}
J.~Gao, B.~Hou, X.~Guo, Z.~Liu, Y.~Zhang, K.~Chen, and J.~Li, ``Secure
  aggregation is insecure: Category inference attack on federated learning,''
  \emph{IEEE Transactions on Dependable and Secure Computing}, 2021.

\bibitem{fu2022label}
C.~Fu, X.~Zhang, S.~Ji, J.~Chen, J.~Wu, S.~Guo, J.~Zhou, A.~X. Liu, and
  T.~Wang, ``Label inference attacks against vertical federated learning,'' in
  \emph{31st USENIX Security Symposium (USENIX Security 22)}, 2022, pp.
  1397--1414.

\bibitem{gong2022coordinated}
X.~Gong, Y.~Chen, H.~Huang, Y.~Liao, S.~Wang, and Q.~Wang, ``Coordinated
  backdoor attacks against federated learning with model-dependent triggers,''
  \emph{IEEE network}, vol.~36, no.~1, pp. 84--90, 2022.

\bibitem{nuding2022data}
F.~Nuding and R.~Mayer, ``Data poisoning in sequential and parallel federated
  learning,'' in \emph{Proceedings of the 2022 ACM on International Workshop on
  Security and Privacy Analytics}, 2022, pp. 24--34.

\bibitem{fang2020local}
M.~Fang, X.~Cao, J.~Jia, and N.~Z. Gong, ``Local model poisoning attacks to
  byzantine-robust federated learning,'' in \emph{Proceedings of the 29th
  USENIX Conference on Security Symposium}, 2020, pp. 1623--1640.

\bibitem{shejwalkar2021manipulating}
V.~Shejwalkar and A.~Houmansadr, ``Manipulating the byzantine: Optimizing model
  poisoning attacks and defenses for federated learning,'' in \emph{NDSS},
  2021.

\bibitem{cao2022mpaf}
X.~Cao and N.~Z. Gong, ``Mpaf: Model poisoning attacks to federated learning
  based on fake clients,'' in \emph{Proceedings of the IEEE/CVF Conference on
  Computer Vision and Pattern Recognition}, 2022, pp. 3396--3404.

\bibitem{chow2021perception}
K.-H. Chow and L.~Liu, ``Perception poisoning attacks in federated learning,''
  in \emph{2021 Third IEEE International Conference on Trust, Privacy and
  Security in Intelligent Systems and Applications (TPS-ISA)}.\hskip 1em plus
  0.5em minus 0.4em\relax IEEE, 2021, pp. 146--155.

\bibitem{zhang2020poisongan}
J.~Zhang, B.~Chen, X.~Cheng, H.~T.~T. Binh, and S.~Yu, ``Poisongan: Generative
  poisoning attacks against federated learning in edge computing systems,''
  \emph{IEEE Internet of Things Journal}, vol.~8, no.~5, pp. 3310--3322, 2020.

\bibitem{zhang2019poisoning}
J.~Zhang, J.~Chen, D.~Wu, B.~Chen, and S.~Yu, ``Poisoning attack in federated
  learning using generative adversarial nets,'' in \emph{2019 18th IEEE
  International Conference On Trust, Security And Privacy In Computing And
  Communications/13th IEEE International Conference On Big Data Science And
  Engineering (TrustCom/BigDataSE)}.\hskip 1em plus 0.5em minus 0.4em\relax
  IEEE, 2019, pp. 374--380.

\bibitem{li2021lomar}
X.~Li, Z.~Qu, S.~Zhao, B.~Tang, Z.~Lu, and Y.~Liu, ``Lomar: A local defense
  against poisoning attack on federated learning,'' \emph{IEEE Transactions on
  Dependable and Secure Computing}, 2021.

\bibitem{qayyum2022making}
A.~Qayyum, M.~U. Janjua, and J.~Qadir, ``Making federated learning robust to
  adversarial attacks by learning data and model association,'' \emph{Computers
  \& Security}, vol. 121, p. 102827, 2022.

\bibitem{zheng2023federated}
J.~Zheng, K.~Li, N.~Mhaisen, W.~Ni, E.~Tovar, and M.~Guizani, ``Federated
  learning for online resource allocation in mobile edge computing: A deep
  reinforcement learning approach,'' in \emph{IEEE Wireless Communications and
  Networking Conference (WCNC)}.\hskip 1em plus 0.5em minus 0.4em\relax IEEE,
  2023, pp. 1--6.

\bibitem{jiang2022model}
Y.~Jiang, S.~Wang, V.~Valls, B.~J. Ko, W.-H. Lee, K.~K. Leung, and
  L.~Tassiulas, ``Model pruning enables efficient federated learning on edge
  devices,'' \emph{IEEE Transactions on Neural Networks and Learning Systems},
  2022.

\bibitem{zheng2022exploring}
J.~Zheng, K.~Li, N.~Mhaisen, W.~Ni, E.~Tovar, and M.~Guizani, ``Exploring deep
  reinforcement learning-assisted federated learning for online resource
  allocation in privacy-preserving edge{I}o{T},'' \emph{IEEE Internet of Things
  Journal}, 2022.

\bibitem{blanchard2017machine}
P.~Blanchard, E.~M. El~Mhamdi, R.~Guerraoui, and J.~Stainer, ``Machine learning
  with adversaries: Byzantine tolerant gradient descent,'' \emph{Advances in
  neural information processing systems}, vol.~30, 2017.

\bibitem{wang2022graphfl}
B.~Wang, A.~Li, M.~Pang, H.~Li, and Y.~Chen, ``Graphfl: A federated learning
  framework for semi-supervised node classification on graphs,'' in \emph{IEEE
  International Conference on Data Mining}.\hskip 1em plus 0.5em minus
  0.4em\relax IEEE, 2022, pp. 498--507.

\bibitem{yuan2022fedstn}
X.~Yuan, J.~Chen, J.~Yang, N.~Zhang, T.~Yang, T.~Han, and A.~Taherkordi,
  ``Fedstn: Graph representation driven federated learning for edge computing
  enabled urban traffic flow prediction,'' \emph{IEEE Transactions on
  Intelligent Transportation Systems}, 2022.

\bibitem{palomar2006tutorial}
D.~P. Palomar and M.~Chiang, ``A tutorial on decomposition methods for network
  utility maximization,'' \emph{IEEE Journal on Selected Areas in
  Communications}, vol.~24, no.~8, pp. 1439--1451, 2006.

\bibitem{cemgil2020autoencoding}
T.~Cemgil, S.~Ghaisas, K.~Dvijotham, S.~Gowal, and P.~Kohli, ``The autoencoding
  variational autoencoder,'' \emph{Advances in Neural Information Processing
  Systems}, vol.~33, pp. 15\,077--15\,087, 2020.

\bibitem{li2024data}
K.~Li, J.~Zheng, X.~Yuan, W.~Ni, O.~B. Akan, and H.~V. Poor, ``Data-agnostic
  model poisoning against federated learning: A graph autoencoder approach,''
  \emph{IEEE Transactions on Information Forensics and Security}, 2024.

\bibitem{li2023exploring}
K.~Li, X.~Yuan, J.~Zheng, W.~Ni, and M.~Guizani, ``Exploring adversarial graph
  autoencoders to manipulate federated learning in the internet of things,'' in
  \emph{International Wireless Communications and Mobile Computing
  (IWCMC)}.\hskip 1em plus 0.5em minus 0.4em\relax IEEE, 2023, pp. 898--903.

\bibitem{wang2020simple}
Y.~Wang, B.~Xu, M.~Kwak, and X.~Zeng, ``A simple training strategy for graph
  autoencoder,'' in \emph{International Conference on Machine Learning and
  Computing}, 2020, pp. 341--345.

\bibitem{zhu2020anomaly}
D.~Zhu, Y.~Ma, and Y.~Liu, ``Anomaly detection with deep graph autoencoders on
  attributed networks,'' in \emph{2020 IEEE Symposium on Computers and
  Communications (ISCC)}.\hskip 1em plus 0.5em minus 0.4em\relax IEEE, 2020,
  pp. 1--6.

\bibitem{hasanzadeh2019semi}
A.~Hasanzadeh, E.~Hajiramezanali, K.~Narayanan, N.~Duffield, M.~Zhou, and
  X.~Qian, ``Semi-implicit graph variational auto-encoders,'' \emph{Advances in
  neural information processing systems}, vol.~32, 2019.

\bibitem{pan2019learning}
S.~Pan, R.~Hu, S.-f. Fung, G.~Long, J.~Jiang, and C.~Zhang, ``Learning graph
  embedding with adversarial training methods,'' \emph{IEEE transactions on
  cybernetics}, vol.~50, no.~6, pp. 2475--2487, 2019.

\bibitem{joyce2011kullback}
J.~M. Joyce, ``Kullback-leibler divergence,'' in \emph{International
  encyclopedia of statistical science}.\hskip 1em plus 0.5em minus 0.4em\relax
  Springer, 2011, pp. 720--722.

\bibitem{molitierno2016applications}
J.~J. Molitierno, \emph{Applications of combinatorial matrix theory to
  Laplacian matrices of graphs}.\hskip 1em plus 0.5em minus 0.4em\relax CRC
  Press, 2016.

\bibitem{lange2010singular}
K.~Lange, ``Singular value decomposition,'' in \emph{Numerical analysis for
  statisticians}.\hskip 1em plus 0.5em minus 0.4em\relax Springer, 2010, pp.
  129--142.

\bibitem{menon2011fast}
A.~K. Menon and C.~Elkan, ``Fast algorithms for approximating the singular
  value decomposition,'' \emph{ACM Transactions on Knowledge Discovery from
  Data (TKDD)}, vol.~5, no.~2, pp. 1--36, 2011.

\bibitem{SHAN2023108950}
\BIBentryALTinterwordspacing
B.~Shan, W.~Ni, X.~Yuan, D.~Yang, X.~Wang, and R.~P. Liu, ``Graph learning from
  band-limited data by graph fourier transform analysis,'' \emph{Signal
  Processing}, vol. 207, p. 108950, 2023. [Online]. Available:
  \url{https://www.sciencedirect.com/science/article/pii/S0165168423000245}
\BIBentrySTDinterwordspacing

\bibitem{boyd2004convex}
S.~P. Boyd and L.~Vandenberghe, \emph{Convex optimization}.\hskip 1em plus
  0.5em minus 0.4em\relax Cambridge university press, 2004.

\bibitem{hebrok2023we}
S.~Hebrok, S.~Nachtigall, M.~Maehren, N.~Erinola, R.~Merget, J.~Somorovsky, and
  J.~Schwenk, ``We really need to talk about session tickets: A
  $\{$Large-Scale$\}$ analysis of cryptographic dangers with $\{$TLS$\}$
  session tickets,'' in \emph{Proceedings of 32nd USENIX Security Symposium
  (USENIX Security 23)}, 2023, pp. 4877--4894.

\bibitem{diaz2019tls}
D.~Diaz-Sanchez, A.~Mar{\'\i}n-Lopez, F.~A. Mendoza, P.~A. Cabarcos, and R.~S.
  Sherratt, ``Tls/pki challenges and certificate pinning techniques for iot and
  m2m secure communications,'' \emph{IEEE Communications Surveys \& Tutorials},
  vol.~21, no.~4, pp. 3502--3531, 2019.

\bibitem{deng2012mnist}
L.~Deng, ``The mnist database of handwritten digit images for machine learning
  research [best of the web],'' \emph{IEEE signal processing magazine},
  vol.~29, no.~6, pp. 141--142, 2012.

\bibitem{xiao2017fashion}
H.~Xiao, K.~Rasul, and R.~Vollgraf, ``Fashion-mnist: a novel image dataset for
  benchmarking machine learning algorithms,'' \emph{arXiv preprint
  arXiv:1708.07747}, 2017.

\bibitem{zheng2024detecting}
J.~Zheng, K.~Li, X.~Yuan, W.~Ni, and E.~Tovar, ``Detecting poisoning attacks on
  federated learning using gradient-weighted class activation mapping,'' in
  \emph{Proceedings of The ACM Web Conference (WWW)}.\hskip 1em plus 0.5em
  minus 0.4em\relax ACM, 2024.

\bibitem{hossain2021desmp}
M.~T. Hossain, S.~Islam, S.~Badsha, and H.~Shen, ``Desmp: Differential
  privacy-exploited stealthy model poisoning attacks in federated learning,''
  in \emph{2021 17th International Conference on Mobility, Sensing and
  Networking (MSN)}.\hskip 1em plus 0.5em minus 0.4em\relax IEEE, 2021, pp.
  167--174.

\bibitem{cao2022flcert}
X.~Cao, Z.~Zhang, J.~Jia, and N.~Z. Gong, ``Flcert: Provably secure federated
  learning against poisoning attacks,'' \emph{IEEE Transactions on Information
  Forensics and Security}, vol.~17, pp. 3691--3705, 2022.

\bibitem{chen2024exploring}
G.~Chen, K.~Li, A.~Abdelmoniem, and L.~You, ``Exploring representational
  similarity analysis to protect federated learning from data poisoning,'' in
  \emph{Proceedings of The ACM Web Conference (WWW)}.\hskip 1em plus 0.5em
  minus 0.4em\relax ACM, 2024.

\end{thebibliography}

\end{document}